# Estimating a new panel MSK dataset for comparative analyses of national absorptive capacity systems, economic growth, and development in low and middle income economies

## Muhammad Salar Khan

### mkhan63@gmu.edu

### Ph.D. Candidate

### Schar School of Policy and Government, George Mason University

**Abstract:**

Within the national innovation system literature, empirical analyses are severely lacking for developing economies. Particularly, the low- and middle-income countries (LMICs) eligible for the World Bank's International Development Association (IDA) support, are rarely part of any empirical discourse on growth, development, and innovation. One major issue hindering panel analyses in LMICs, and thus them being subject to any empirical discussion, is the lack of complete data availability. This work offers a new complete panel dataset with no missing values for LMICs eligible for IDA's support. I use a standard, widely respected multiple imputation technique (specifically, *Predictive Mean Matching*) developed by Rubin (1987). This technique respects the structure of multivariate continuous panel data at the country level. I employ this technique to create a large dataset consisting of many variables drawn from publicly available established sources. These variables, in turn, capture six crucial country-level capacities: technological capacity, financial capacity, human capital capacity, infrastructural capacity, public policy capacity, and social capacity. Such capacities are part and parcel of the *National Absorptive Capacity Systems* (NACS). The dataset—MSK dataset—thus produced contains data on 47 variables for 82 LMICs between 2005 and 2019. The dataset has passed a quality and reliability check and can thus be used for comparative analyses of national absorptive capacities and development, transition, and convergence analyses among LMICs.

## Introduction

*Without data, you're just another person with an opinion. (William Edwards Deming)*

The National Innovation System (NIS) focuses on a broad range of variables, activities, institutions, and their interactions that can foster economic growth and development in countries (Edquist 2006). However, this literature underrepresents the global South. One of the major problems for this lack of reasonable representation stems from the lack of data for low- and middle-income countries (LMICs). By resulting in the exclusion of LMICs in empirical analyses, missing data lead to either positively or negatively biased results that manifest themselves in over and underestimated effect sizes.

Despite the general limitations, several studies have recently investigated NIS and its relationship with growth and development in some developing economies (Choi and Zo 2019; Intarakumnerd, Chairatana, and Tangchitpiboon 2002; Lundvall et al. 2009; Casadella and Uzunidis 2017). Other studies, using capacities as a way to operationalize NIS, have employed available data for diverse samples of countries to estimate the quantitative impact of financial, technological, and social capacities of countries on their economic growth and development process (Khayyat and Lee 2015; Fagerberg and Srholec 2008; 2017; Archibugi and Coco 2005; Gebauer, Worch, and Truffer 2012; Andersson and Palacio Chaverra 2017).

Inspired by the studies on capacities and economic development, I rigorously operationalize a thorough list of capacities that capture innovation, knowledge absorption, and learning processes in LMICs and include those capacities in a formal framework of National Absorptive Capacity System (NACS) that I propose (Khan 2021). A firm-level concept of "absorptive capacity," as advanced by Cohen and Levinthal (1990), particularly motivates the NACS framework. As a modified version of NIS, NACS considers an LMIC an "economic learning" entity that absorbs, creates and deploys knowledge, learning, and skills subject to the strength of its local capacities. To study NACS and its evolution in LMICs and to further examine the impact of the framework capacities on economic development in LMICs, complete panel data (country-year observations) on variables that measure capacities is required. Unfortunately, such variables are not wholly available across LMICs eligible for the World Bank's International Development Association (IDA) support and foci of this study.[1] Hence there is a dire need to fix this problem of missing data for those LMICs, presumably prime candidates for development, learning, economic growth, and innovation. Therefore, in this article, I build a full, recent, and new dataset on variables constituting capacities within LMICs, using statistical and machine learning established tools.

Data incompleteness, commonly called the missing data problem, severely hampers empirical research. Various research fields have extensively investigated missing data dynamics, their consequences, and possible remedies (Nugroho and Surendro 2019; Xue et al. 2017; Gilbert and Sonthalia 2018; Enders 2017a; Ginkel et al. 2020; Jones and Tonetti 2020). However, fields as nascent as NIS and NACS have yet to thoroughly investigate the nuances, processes, and implications of missing data. One significant repercussion of missing data is that the current empirical literature on NIS and economic growth suffers from an imbalance. The literature either focuses on many countries with a limited span of time (Fagerberg and Srholec 2008) or analyzes a few economies with an extended time (Castellacci and Natera 2016; Erdal and Göçer 2015). The former strand of literature can only provide a limited study of the evolution within NIS and NACS, whereas the latter strand prevents analyses in many LMICs. Hence neither is ideal; while the former is static, the latter is not representative of the LMICs.

This article systematically compiles, estimates, and imputes an incomplete dataset to alleviate the missing data problem in LMICs eligible for IDA support. It employs multiple imputation (MI) approach that *efficiently* and *consistently* estimates missing data and generates a panel dataset for 82 LMICs between 2005 and 2019. MI uses state-of-the-art statistical methods to address the missing data problem (Rubin 1996; Enders 2017b). Many research fields have embraced such techniques (Miok et al. 2019; Nissen, Donatello, and Van Dusen 2019; Gondara and Wang 2018; Pedersen et al. 2017). This work explicitly employs a variable-by variable (sequential or chained)

---

[1] Eligibility for IDA support depends mainly on a country's relative poverty. Relative poverty is defined as GNI per capita below an established threshold, and it is updated annually ($1, 185 in the fiscal year 2021). IDA also supports some countries, including several small island economies, that are above the operational cutoff but lack the creditworthiness needed to borrow from the International Bank for Reconstruction and Development (IBRD). Some countries, such as Nigeria and Pakistan, are IDA-eligible based on per capita income levels and are also creditworthy for some IBRD borrowing. They are termed as "blend" countries. http://ida.worldbank.org/about/borrowing-countries

Since IDA eligibility is based off GNI per capita, countries graduate and reinter (reverse graduate in the list). I have data on 82 countries (74 among them are still eligible for IDA resources and 8 countries recently graduated). For a list of IDA graduates, please check: http://ida.worldbank.org/about/ida-graduates

predictive mean matching (PMM) technique (Santos and Conde 2020). As an MI conditional modeling approach, PMM imputes missingness dependent on observed data in continuous, panel variables that do not have to be normally distributed (Santos and Conde 2020; Morris, White, and Royston 2014; Akmam et al. 2019). This technique returns meaningful imputations that respect the data distribution of the original incomplete dataset (observed dataset).

Castellacci and Natera (2011) conducted a similar data compilation study (CANA hereon). The researchers estimate a CANA dataset for 134 countries between 1980 and 2008 using an MI algorithm developed by Honaker and King (2010). The MSK dataset is similar to CANA dataset as both are panel datasets estimated using novel M1 techniques. Similarly, both datasets have a roughly identical structural build of NACS and NIS. For instance, they contend that such systems are measured by dimensions (CANA) and capacities (MSK), which, in turn, are captured by many variables interacting in multiple ways. Although this article builds on CANA, it is different in several ways. First, as opposed to the CANA dataset, the MSK dataset estimated here focuses on relatively more data-deficient and economically poor IDA-eligible countries. Secondly, though the MSK dataset employs some of the CANA dataset variables, it has an entirely different functional and operational conception of the capacities and the variables used to operationalize those capacities. Particularly, Public Policy and Social Capacity are operationalized very differently. Additionally, the MSK dataset includes an extended set of other relevant variables to measure capacities (MSK consists of 47 variables for all economies in the dataset, whereas CANA consists of 34 variables for all economies and another seven variables for a restricted set of countries within the dataset). Third, the timeframe for this study is truncated to fifteen years, not only because it is a decent period for panel analysis but also because of pragmatic concerns regarding data availability, particularly on public and social policy capacity variables. The World Bank Group's country offices started collecting these variables in the IDA-eligible countries from 2005 onwards ("Country Policy and Institutional Assessment" 2014).

The last vital distinction worth considering is that the CANA dataset is estimated using Honaker and King's (2010) Expectation-Maximization algorithm. The MSK, on the other hand, is estimated using the Multiple Imputation by Chained Equations Predictive Mean Matching (*MICE PMM*) algorithm. Although the *EM* algorithm is efficient and undoubtedly suitable for panel data, it forces a normal distribution on the imputed data regardless of the distribution structure (skewed, unimodal, bimodal) in the observed data (Shireman, Steinley, and Brusco 2017). In contrast, the MICE PMM algorithm preserves the distribution pattern of observed data in the imputed values (Vink et al. 2014), and it has been used for panel data imputation (Kleinke 2017). Besides preserving the distribution pattern in the imputed values, the MICE PMM is best suited for this study because the data structure is heteroskedastic, and associations among variables are nonlinear.[2]

In short, this article contributes to the literature by constructing a complete dataset and establishing its relevance for panel analyses of NACS and economic growth, among other analyses, in LMICs. A standard MICE PMM algorithm is employed to construct this dataset. The panel dataset, hence obtained, is complete, i.e., it has no missing values. It consists of 47 variables grouped into six vital capacities for each country: technological capacity, financial capacity, human capital capacity, infrastructural capacity, public policy capacity, and social capacity. The incomplete (original or observed) dataset is constructed from reputable data sources and contains many missing values. The MSK dataset is estimated from this observed dataset, which provides information on 82 LMICs between 2005 and 2019 (total observations are 1,230 country-year observations). A four-way quality check establishes this dataset's reliability and usefulness for researchers interested in panel analyses of absorptive capacity and innovation system, economic development, economic policy, and convergence analysis within LMICs.

The rest of the paper is shaped as follows. Section 2 gives a brief literature landscape, the association between NIS and NACS, and discusses the missing data and its implications on methodologies. Section 3 further discusses the importance of handling missing data, strategies to address missingness, and underlying missing data mechanisms. Section 4 elaborates on Multiple Imputation and MICE PMM technique. Section 5 discusses the MSK dataset and the steps taken to develop this dataset. Section 6 carries out a brief descriptive analysis of the MSK dataset, and Section 7 conducts a quality check of the estimated dataset. Lastly, Section 8 concludes by

---

[2] For heteroskedasticity, I checked for variances of the variables in the data. Most of them differed. For instance, variance of *days to enforce contract* is 80 times larger than the variance of *days to start business*. Similarly, I looked at scatter plots for the variables, which showed funnel shaped spread for many variables.

For associations among variables, I looked at scatterplots again. They showed non-linear relationships.

summarizing the results and implications of this work. The Appendix includes graphs and tables, conveying more information on how the database is constructed and other dataset characteristics.

## 2. From NIS to NACS: Comparative analyses of national systems and growth, and development and the problem of missing data in developing economies

The concept of NIS emerged in the 1990s (Nelson 1993; Freeman 1995; Edquist 1997). It considers systems, activities, institutions, and interactions as the driving force behind economic growth and development (Edquist 2006; López-Rubio, Roig-Tierno, and Mas-Verdú 2021). The strength of these factors explains cross-country differences in growth, development, and innovation. Around the time NIS emerged, Cohen and Levinthal (1990) developed the idea of "absorptive capacity" to explain how learning is consolidated in a firm and how it impacts a firm's growth. In the early 2000s, researchers extended the firm-level concept to a national level (Narula 2004; Criscuolo and Narula 2008). They developed a theoretical framework for aggregating national absorptive capacities upwards from a firm level. Other empirical studies also applied the idea in a national setting (Fagerberg and Srholec 2017). These works used different capacities emerging in NIS literature (such as technological and social capacities) as proxies for national absorptive capacity. In this essence, NACS is essentially an offshoot of NIS.

Earlier, foundational theoretical and empirical work on NIS focused mainly on prosperous economies (Nelson 1993; Edquist 2001). Later, NIS literature theoretically included developing countries, as they considered developing countries "national economic learning" entities and "imitation" centers (Viotti 2002; Lundvall et al. 2009; Fagerberg and Verspagen 2002). National level capacities literature examining the impact of capacities on economic development also included some developing economies in their analyses (Fagerberg and Srholec 2017). However, because of the lack of data in LMICs, such studies had to compromise operationalizing the complex and multifaceted capacities proposed in NIS and NACS. Similarly, the lack of data on many vital variables perhaps trimmed the list of essential capacities in their analyses.

Another critical challenge that missing data posed is that it limited the application of studies methodologies in many LMICs. In general, quantitative studies of capacities and development used mainly two different methodologies: panel regression analyses and composite indicator analyses.

Panel regression analyses examine the empirical relationship between a few capacity variables and comparative national differences in GDP per capita growth across countries (Teixeira and Queirós 2016; Ali, Egbetokun, and Memon 2018). While powerful as they consider the dynamic nature of capacities, such panel studies either ignore or drop off many LMICs because longitudinal data for many variables are missing in these countries. As a result, the coefficients of interest obtained through panel analyses do not provide information about the economically poor economies. Using econometric terminology, the estimates from such studies exhibit an upward or downward bias by overestimating or underestimating the effect of *capacities* on *economic growth*.

On the other hand, composite indicator analyses establish a country's comparative standing against other countries by building aggregate or composite indicators that denote different dimensions of technological and social capabilities (Fagerberg and Srholec 2008; 2015). As opposed to panel analyses, the composite analyses consider many countries, including some LMICs. However, since most LMICs have limited data, such studies are usually static (one-year studies), ignoring how NACS evolved. Also, not all LMICs have data on all the variables of interest available for one particular year. Therefore, even composite analyses cannot possibly include all LMICs.

Generally, data availability restricts the number of countries and periods used in the analyses. Both methodologies are challenging for developing countries, particularly for LMICs eligible for IDA resources, which are the focus of this study. This article contributes to alleviating the problems stemming from missingness by constructing a new complete panel dataset. A statistical technique called MICE PMM is employed to estimate the missing values in the original incomplete data sources (Rubin 1996). Out of many imputation suites, this article considers MICE PMM because they are powerful, efficient, consistent, convenient, and reliable. The following section elaborates on why it is essential to adequately handle missing data and what strategies could be used to deal with missing data.

### 3. Properly Handling Missing Data- why it is crucial, mechanisms underlying missing data, and strategies to handle missing data

It is essential to consider the missing data problem carefully to obtain accurate estimates of the parameters of interest in any analysis. Missing data pose many dilemmas in data analysis. The chief dilemma is, if a researcher uses original data by excluding subjects with missing data from the study, the researcher will not use all the existing information in the data, most likely causing over-or underestimated parameters (aka 'biased parameters'). To treat biasedness in parameters (overestimation and underestimation) caused due to the exclusion of subjects in the analysis, a researcher can impute the missing data. During the imputation process, however, the researcher should take utmost care in preserving variability found in existing data and incorporating uncertainty underlying any missing data. Therefore, it is imperative to employ proper and standard imputation methodologies to estimate a reliable dataset.

Provided that the imputation technique is sound, one may get reliable imputations. The first step in getting the imputation technique right essentially means being very mindful of the missing data pattern and what might have caused it. The literature considers three potential mechanisms underlying missing data (Papageorgiou et al. 2018).

***Missing Completely At Random (MCAR)-*** Missing is MCAR if it is genuinely by chance, i.e., missingness is independent of data characteristics. In other words, missingness in MCAR is not related to any nonmissing or missing values in the data set. For example, the random loss of a blood sample in the lab suggests MCAR.

***Missing At Random (MAR)-*** Data exhibits MAR if the missingness is due to observed but not unobserved data. In other words, the observed data explains the missingness. For example, women may be less likely to report their age, regardless of their actual age.

***Missing Not At Random (MNAR)-*** In such a mechanism, missing values explain missingness. For example, individuals with higher salaries may be less willing to answer survey questions about their pay. Another example of MNAR relates to a person not attending a drug test because they took drugs the night before.

To properly handle data, understanding the abovementioned mechanisms underlying missing data is extremely important. If a researcher fails to understand the missing data pattern and the underlying mechanism and imputes missing values, the missing data may be mistreated. Consequently, resulting results will exhibit insufficient statistical power, upward or downward biases in parameters of interest, under or overestimated standard errors of the parameters, and other inaccurate findings.

Two main strategies are employed to handle missing data: 1) deletion and 2) substitution and imputation (Cook 2021). Deletion (also called complete or available-case analysis) is of two kinds:

*pairwise* or *listwise* deletion (Lang and Little 2018). Both these kinds exclude observations with missing values while analyzing data (Lang and Little 2018). Imputation or substitution imputes or substitutes for missing values, and it is also of two main types: single imputation and multiple imputation (Ginkel et al. 2020). Single Imputation produces one complete dataset when imputing for missing values. It can be accomplished via several techniques such as mean substitution, mode substitution, nearest neighbor-based imputation, regression, or cold deck imputation (Silva-Ramírez, Pino-Mejías, and López-Coello 2015). Multiple Imputation (MI), on the other hand, produces multiple imputed data sets, employs a statistical analysis model to each one, and eventually merges all analysis results to generate an overall result (Enders 2017b). Based on various data pattern assumptions and underlying data structures, MI is executed in many ways, such as parametric (*Multivariate Normal MI*) or semiparametric approaches (*Multiple Imputation by Chained Equations* including *Predictive Mean Matching*).[3] Another imputation technique, performed in one or many runs, is Expectation-Maximization (EM) algorithm. EM is an iterative algorithm that finds maximum likelihood estimates in parametric models (Honaker, King, and Blackwell 2011). These strategies have both pros and cons (see Appendix A). Of those strategies, this article employs Multiple Imputation by Chained Equations (MICE), specifically Predictive Mean Matching (PMM), for imputing missing values that do not observe a normal distribution. MICE PMM is not only a convenient, standard, and reliable technique, but it also gives very accurate and plausible estimates for the data under consideration (Kleinke 2017; K.-H. Kim and Kim 2020). The next section briefly describes MI, MICE, and PMM.

## 4. The Multiple Imputation Method and Predictive Mean Matching

Rubin (1987) first introduced multiple imputation methodology as an *efficient* statistical methodology to estimate missing values in a dataset. Several other researchers also explain this technique (Little and Rubin 1989; Rubin and Schenker 1986). This methodology has evolved into various methods over the years, catering to missingness in diverse data models. MI overcomes many of the problems associated with deletion and other single imputation techniques (Shi et al.

---

[3] Parametric models are statistical models that have a finite number of parameters. Parametric modeling creates a model for known facts (parameters) about population. An example is normal distribution model with parameters mean and standard deviation. In general, parametric models work well with normally distributed data. On the other hand, nonparametric models have infinite number of parameters and they relax normality assumption. Usually, they assume that the data is not normally distributed. Semiparametric models have both parametric (finite-dimensional i.e., it is easy to research and understand) and nonparametric (i.e., beyond the range of ordinary statistical methods) components. Semiparametric also relaxes normality assumption. More details, see Pace (1995).

2020; Afghari et al. 2019). In addition, the methodology returns efficient and accurate estimates and preserves *variability,* which is otherwise lost using other single imputation techniques (such as mean or cold deck imputation).

MI is valid under MAR (*Missing at Random*) assumption (Afghari et al. 2019). Therefore, MI estimates missing values by using available, observed data (Harel et al. 2018).

Since there is uncertainty about missing data values, the estimation process is repeated *m* times (this step refers to the *imputation stage*). From the imputation stage, *m* complete datasets are generated. In the next stage (*analysis stage*), econometric analyses of interest are separately performed on *m* datasets. Finally, all these multiple results are combined (pooled) to obtain a final value of the coefficient of interest, for instance, regression coefficients (*pooling stage*). In short, a standard MI process produces multiple imputed datasets, applies a statistical analysis model to each dataset, and then integrates all analysis results to create an overall result (see Figure 1 below).

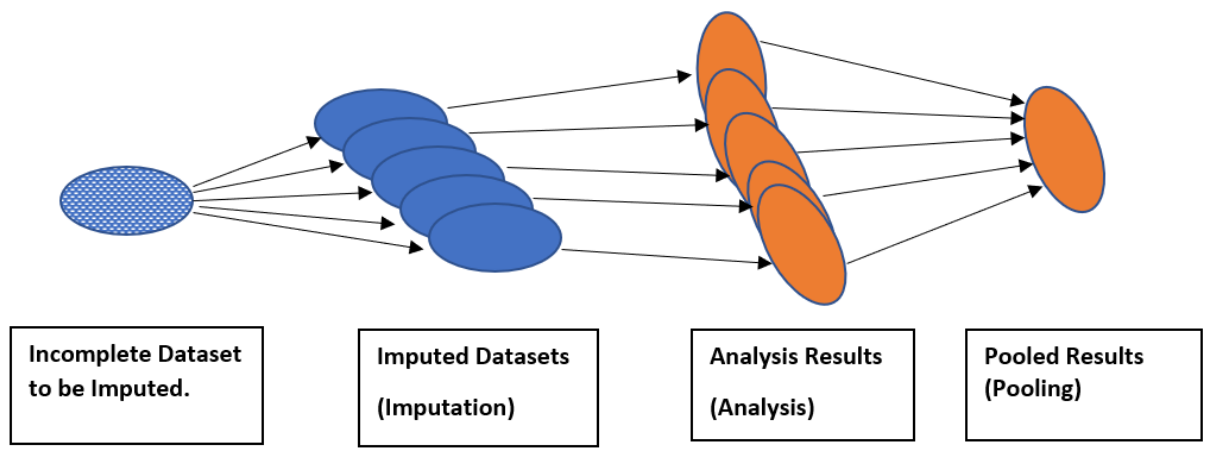

Figure 1: Shows a standard Multiple Imputation process. In the first step (imputation stage), missing data at hand, shown in white dots, are imputed (all in blue now showing imputation happened) to create *m* imputed datasets. Following imputation, each imputed dataset is separately analyzed using standard methods (such as OLS regression). Lastly, the analysis results are combined using Rubin's rules (1987).

Suppose the imputation model at the imputation stage is specified correctly and the data exhibit a normal distribution. In that case, MI yields consistent parameter estimation and confidence intervals that incorporate uncertainty because of the missing data (Morris, White, and Royston 2014). To clarify, the correct specification of an imputation model entails the inclusion of variables considered to predict missingness and variables associated with the variable being imputed and the outcome variable of the analysis model (Morris, White, and Royston 2014; Kg et al. 2006).

MI consists of many *parametric* (satisfy multivariate normality assumption) and *semiparametric* (works even if the multivariate normality assumption is violated) approaches. One of the common parametric approaches is Multivariate Normal distribution MI (*MVN*). This approach assumes all imputed variables to follow a joint multivariate normal distribution. Conversely, MI by Chained Equations (MICE) is a semiparametric approach that does not assume a joint MVN distribution but instead considers a different distribution for each imputed variable (Zhang 2016b). Unlike MVN, MICE employs a sequential (variable-by-variable) approach while incorporating functional relationships among variables and data characteristics such as ranges. Within MICE, one can either use Linear Regression or Predictive Mean Matching (PMM) for continuous variables. This article carries out the PMM technique to impute missing values. PMM relaxes most of the assumptions of parametric MI techniques (Akmam et al., 2019). Hence, it is handy for imputing quantitative variables that are not normally distributed (Lee and Carlin 2017). In the PMM, the missing value for an observation (considered as a 'recipient') is imputed by the observed value from another observation (called as 'donor') with a similar predicted mean outcome as follows (Akmam et al. 2019; Luo and Paal 2021):

In the *imputation* stage, for every missing value, the PMM algorithm structures a small set of donors (typically 5 or 10) from all complete cases that have *predicted* values closest to the predicted value for the missing value. Next, one donor is randomly drawn from the neighborhood pool. The observed value of such a donor is assigned to the missing value. This procedure is conducted *m* times, which generates *m* datasets. After the imputation stage, *analysis* and *pooling* stages follow the same pattern as any standard MI. Like any MI, in the analysis stage *m* times analyses are conducted, and in the pooling stage, these results are combined to get a single estimate.

A more step-by-step computational process within the imputation stage of PMM is explained below:

Suppose there is a variable (X) that has missing values and another set of variables (Vs) to be used to impute X, the software (STATA or R) carries out the following computations in the imputation stage:

1- Firstly, it estimates a linear regression of X on Vs for complete observations (those with no missing values). This step produces a set of coefficients *a*.

2- Secondly, it randomly draws from the "posterior predictive distribution" of *a*.[4] This step generates a new set of coefficients *a\**. (this step ensures variability in the imputed values produced later on).

3- Thirdly, the software uses coefficients *a\** to generate predicted values for X for all observations.

4- Fourthly, for each observation with a missing value of X, the software identifies a set of observations with observed X (called donors or neighbors) whose predicted values are roughly close or similar to the predicted value for the observation with missing data.

5- Lastly, from the neighborhood pool identified, it randomly chooses one donor and designates its observed value to fill in for the missing value.

For each completed dataset, steps 2 through 5 are conducted. The key idea is constructing a right donor pool from where observations with missing data will be matched with observations with available data (Allison 2015). Researchers have answered how many donors or neighbors should be in the donor pool (Morris, White, and Royston 2014; Allison 2015). They assert that the size of the pool depends on sample size. In general, for most situations, these studies suggest k=10 or k=5. The default in the Stata MI command is k=1.

In short, PMM is simple to perform and a versatile method. It relaxes normality distribution assumption, which is not always observed in continuous data. Since PMM imputations are based on observed values in the neighborhood, therefore they are much more realistic. Unlike other techniques such as EM or MVN, PMM does not produce imputations outside the observed values; thus, they overcome the problems with meaningless imputations. Compared to other suites such as Normal Linear Regression imputation, PMM is also less susceptible to model specification, and it can handle many variables irrespective of their distributions (Kleinke 2017). While imputing from the neighboring donor candidates, it incorporates nonlinearities (nonlinear associations

---

[4] The posterior predictive distribution is the distribution of possible unobserved values conditional on the observed values (Williams et al. 2020)

among variables) and returns the same distribution for missing data present in the observed data (Kleinke 2017).

## 5. A New MSK Panel Dataset

Here I am presenting the main features of the MSK dataset. The dataset has been compiled and estimated after applying the MICE predictive mean matching technique described in the previous section. The complete dataset consists of information for many pertinent variables and for all LMICs eligible for IDA support over time (panel data). Specifically, the dataset contains complete data for 47 variables for 82 countries between 2005 and 2019 (1,230 country-year observations).

This new complete dataset offers ample statistical content to conduct longitudinal comparative country analyses of national absorptive capacity systems (NACS) within LMICs. Among other valuable insights, such analyses illustrate the relative standing of LMICs. Similarly, the dataset's time-series feature enlightens how LMICs' NACS evolved in the last one and a half decades. Immediate use of the dataset would entail estimating the relationship between the variables within the dataset (capacities constituting NACS) and the LMICs' economic development. Such an exercise will offer crucial lessons on economic growth and development to leading and lagging LMICs. Similarly, another use will involve clustering LMICs into different groups based on capacities scores.

Since NACS are multifaceted, any analysis of NACS would involve a large number of possibly relevant variables interacting in many ways. Therefore, the MSK dataset embraces a multidimensional operationalization of NACS. In this dataset, the NACS constitutes six capacities drawn from the literature. In addition, various *incoming* flows from abroad (learning, knowledge, skills, and technology) also may influence the NACS. Figure 2 represents these capacities of NACS while alluding to the incoming flows. The six capacities are: 1) Technological capacity, 2) Financial capacity, 3) Human capacity, 4) Infrastructural capacity, 5) Public Policy capacity, and 6) Social capacity. The discussion of these capacities (and incoming flows) and how encompassing they are compared to other narrow definitions of capacities is beyond this article's scope (please see Khan 2021 for this discussion). However, the central hypothesized idea behind this dataset's construction is that LMICs that are severely lacking in data need to appreciate that these capacities and their dynamic interaction drive economic development and science, technology, and innovation (STI) in those economies. For this purpose, development economists and STI policymakers need to have access to panel statistical data (country-year observations) on these capacities, which would help them conduct empirical analyses.

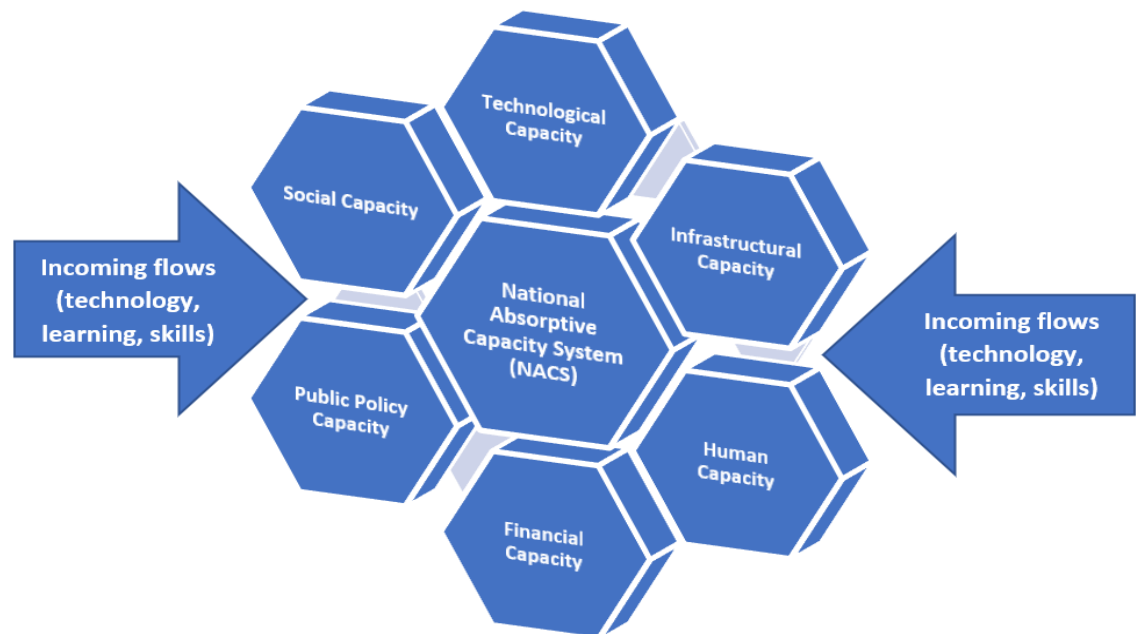

Figure 2: Shows National Absorptive Capacity System (NACS) and its capacities. These six capacities constitute NACS. Incoming flows mediate capacities within NACS.

Literature on NIS helped identify 64 variables, likely constituting one of these capacities in NACS. After performing imputation analysis, the list of variables was reduced. Resultantly, the MSK dataset consists of 47 variables, as shown in Table 1. As a matter of good practice, the table also compares descriptive statistics (mean, standard deviation, minimum, maximum, and observation count) of the variables in the new (complete) dataset with descriptive statistics for corresponding variables in the observed (incomplete) dataset. The last column of the table reports the share of missing data present in the original dataset. As can be seen, the missingness is very high for some variables; missingness ranges from 0.89% to about 87%. A quick look at the table shows that descriptive statistics of the two data (complete and incomplete) do not differ much. This is one of the many ways to show that the complete dataset is sufficiently reliable (this will be elaborated in the forthcoming section).

**Table 1: Descriptive Statistics of New MSK Dataset Vs. Incomplete Observed Dataset (for more details on the variables, please consult Appendix B)**

| Capacity and Variables | Variable code | MSK Dataset | | | | | Observed Dataset | | | | | Missing% |
|---|---|---|---|---|---|---|---|---|---|---|---|---|
| | | Obs | Mean | Std. Dev. | Min | Max | Obs | Mean | Std. Dev. | Min | Max | |
| **TECHNOLOGY CAPACITY** | | | | | | | | | | | | |
| Sci & tech. articles | tscitjar | 1230 | 1270.77 | 9395.79 | 0 | 135787.8 | 1,148 | 1236.60 | 9247.52 | 0 | 135787.8 | 6.67% |
| Intellectual payments (mil) | tippay | 1230 | 65.35 | 492.20 | -13.92 | 7906 | 818 | 87.80 | 601 | -13.97 | 7909 | 33.50% |
| Voc. & tech. students (mil) | tsecedvoc | 1230 | 111698.6 | 253483.79 | 0 | 2300769 | 571 | 121436.2 | 277829.5 | 0 | 2300769 | 53.58% |
| R&D expend. % of GDP | trandd | 1230 | .21 | .16 | .01 | .86 | 225 | 0.25 | 0.19 | 0.01 | 0.859 | 81.71% |
| R&D researchers (per mil) | tresinrandd | 1230 | 162.65 | 225.9 | 5.94 | 1463.77 | 148 | 256 | 317 | 5.93 | 1463.77 | 87.97% |
| R&D technicians (per mil) | ttechinrandd | 1230 | 57.02 | 63.01 | .13 | 627.73 | 144 | 55.27 | 70.22 | 0.13 | 627.73 | 88.29% |
| High-tech exports (mil) | thigexperofmanex | 1230 | 6.23 | 9.29 | 0 | 68.14 | 547 | 5.80 | 8.74 | 0.00008 | 68.14 | 55.53% |
| ECI (econ. complexity) | teciscore | 1230 | -.72 | .63 | -3.04 | .82 | 892 | -0.77 | 0.62 | -3.04 | 0.82 | 27.48% |
| **FINANCIAL CAPACITY** | | | | | | | | | | | | |
| Tax revenue (% of GDP) | ftaxrpergdp | 1230 | 16.22 | 11.71 | 0 | 149.28 | 583 | 15.7 | 11 | 0.0001 | 149.28 | 52.60% |
| Business startup cost | fcosbstpropergni | 1230 | 85.38 | 137.76 | 0 | 1314.6 | 1,154 | 79 | 120.2 | 0 | 1314.6 | 6.18% |
| Domestic credit by banks | fdomcrprsecbybkpergdp | 1230 | 25.07 | 20.37 | .5 | 137.91 | 1,100 | 26.3 | 20.85 | 0.5 | 137.91 | 10.57% |
| Days to start business | ftdaystobusi | 1230 | 35.34 | 37.71 | 1 | 260.5 | 1,154 | 34.48 | 36.45 | 1 | 260.5 | 6.18% |
| Days enforcing contract | fdaystoenfctt | 1230 | 666.61 | 329.52 | 225 | 1800 | 1,154 | 662.2 | 322.4 | 225 | 1800 | 6.18% |
| Days to register property | fdaystoregpro | 1230 | 87.33 | 97.58 | 1 | 690 | 1,104 | 81 | 89.6 | 1 | 690 | 10.24% |
| Openness measure | fopenind | 1230 | .11 | .08 | .01 | .44 | 847 | 0.11 | 0.08 | 0.009 | 0.44 | 31.14% |
| Days to electric meter | fdaystoobteleconn | 1230 | 37.24 | 33.64 | 2.5 | 194.3 | 153 | 34.3 | 31.31 | 2.5 | 194.3 | 87.56% |
| Business density | fnewbusdenper1k | 1230 | 1.06 | 1.47 | .01 | 12.31 | 583 | 1.19 | 1.67 | 0.006 | 12.30 | 52.60% |
| Financial accountholders | faccownperofpop15p | 1230 | 30.94 | 22.53 | 1.52 | 92.97 | 160 | 30 | 19.28 | 1.52 | 92.97 | 86.99% |
| Commercial banks | fcombkbr1k | 1230 | 10.49 | 11.99 | .27 | 71.23 | 1,099 | 10.58 | 12.045 | 0.27 | 71.23 | 10.65% |
| **HUMAN CAPITAL CAPACITY** | | | | | | | | | | | | |
| Primary enrollment (gross) | hprimenrollpergross | 1230 | 103.36 | 18.18 | 23.36 | 149.96 | 911 | 103.4 | 18.15 | 23.36 | 149.95 | 25.93% |
| Sec. enrollment (gross) | hsecenrollpergross | 1230 | 57.49 | 25.99 | 5.93 | 123.03 | 711 | 58.03 | 26.63 | 5.93 | 123.03 | 42.20% |
| Primary pupil-teacher ratio | hpupteapriratio | 1230 | 34.43 | 14.36 | 8.68 | 100.24 | 751 | 35.3 | 14.63 | 8.68 | 100.24 | 38.94% |
| Primary completion rate | hprimcompra | 1230 | 79.41 | 20.89 | 26.1 | 134.54 | 735 | 78.83 | 20.72 | 26.09 | 134.54 | 40.24% |
| Govt. expend. on educ. | hgvtexpedupergdp | 1230 | 4.36 | 2.22 | .69 | 12.9 | 615 | 4.06 | 1.91 | 0.69 | 12.90 | 50% |
| Human Capital Index 0-1 | hhciscale0to1 | 1230 | .42 | .09 | .29 | .69 | 154 | 0.43 | 0.09 | 0.28 | 0.69 | 87.42% |
| Advanced educ. labor | hlfwithadedu | 1230 | 75.5 | 10.55 | 39.97 | 96.36 | 265 | 76.08 | 10.29 | 40 | 96.36 | 78.46% |
| Compulsory educ. (years) | hcompeduyears | 1230 | 8.45 | 2.16 | 4 | 15 | 1,028 | 8.57 | 2.16 | 4 | 15 | 16.42% |
| Industry employment | hempinduspertotem | 1230 | 14.52 | 7 | .64 | 32.59 | 1,125 | 14.08 | 6.94 | 0.64 | 32.59 | 8.54% |
| Service employment | hempserpertotem | 1230 | 39.43 | 15.05 | 7.16 | 75.34 | 1,125 | 37.8 | 14.24 | 7.16 | 75.34 | 8.54% |

| | | | MSK Dataset | | | | | Observed Dataset | | | | |
|---|---|---|---|---|---|---|---|---|---|---|---|---|
| **INFRASTRUCTURE CAPACITY** | | | | | | | | | | | | |
| Mobile subscriptions | imobsubper100 | 1230 | 59.12 | 38.15 | .26 | 181.33 | 1,219 | 59.19 | 38.17 | 0.26 | 181.33 | 0.89% |
| Access to electricity | iaccesselecperpop | 1230 | 57.02 | 31.3 | 1.24 | 100 | 1,135 | 56.77 | 31.32 | 1.24 | 100 | 7.72% |
| Broadband subscriptions | ibdbandsubper100 | 1230 | 1.97 | 4.12 | 0 | 25.41 | 1,114 | 2.02 | 4.23 | 0 | 25.41 | 9.43% |
| Telephone subscriptions | itelesubper100 | 1230 | 5.31 | 7.39 | 0 | 32.85 | 1,218 | 5.29 | 7.40 | 0 | 32.85 | 0.98% |
| Energy use (per capita) | ienergyusepercap | 1230 | 560.21 | 392.9 | 9.55 | 2246.92 | 471 | 553 | 376.25 | 9.54 | 2246.92 | 61.71% |
| Logistic perf. Index 1-5 | ilpiquoftratraninfr | 1230 | 2.18 | .33 | 1.1 | 3.34 | 372 | 2.19 | 0.32 | 1.1 | 3.34 | 69.76% |
| Internet users | iindintperpop | 1230 | 16 | 16.3 | .03 | 89.44 | 1,209 | 16 | 16.33 | 0.031 | 89.44 | 1.71% |
| **PUBLIC POLICY CAPACITY** | | | | | | | | | | | | |
| CPIA econ. mgmt. | pcpiaeconmgtcl1to6 | 1230 | 3.39 | .69 | 1 | 5.5 | 1,132 | 3.40 | 0.67 | 1 | 5.5 | 7.97% |
| Public sect. mgmt. & instit | pcpiapsmgandinscl1to6 | 1230 | 3.06 | .5 | 1.4 | 4.2 | 1,132 | 3.06 | 0.48 | 1.4 | 4.2 | 7.97% |
| Sructural policies | pcpiastpolclavg1to6 | 1230 | 3.3 | .54 | 1.17 | 5 | 1,132 | 3.31 | 0.52 | 1.17 | 5 | 7.97% |
| Statistical capacity 0-100 | pscapscoravg | 1230 | 59.82 | 14.89 | 20 | 96.67 | 1,206 | 59.9 | 14.87 | 20 | 96.67 | 1.95% |
| Legal Rights Index 0-12 | pstrengthoflegalright | 1230 | 4.83 | 3.1 | 0 | 11 | 565 | 5.27 | 3.05 | 0 | 11 | 54.07% |
| **SOCIAL CAPACITY** | | | | | | | | | | | | |
| Human resources rating | scpiabdhmanres1to6 | 1230 | 3.52 | .63 | 1 | 4.5 | 1,132 | 3.52 | 0.61 | 1 | 4.5 | 7.97% |
| Equity of public resc use | scpiaeqofpbresuse1to6 | 1230 | 3.38 | .64 | 1 | 4.5 | 1,132 | 3.39 | 0.62 | 1 | 4.5 | 7.97% |
| Social protection rating | scpiasocprorat1to6 | 1230 | 3.03 | .59 | 1 | 4.5 | 1,128 | 3.04 | 0.58 | 1 | 4.5 | 8.29% |
| Social inclusion o.. | scpiapolsocinclcl1to6 | 1230 | 3.28 | .51 | 1.5 | 4.3 | 1,129 | 3.28 | 0.50 | 1.5 | 4.3 | 8.29% |
| National headcount poverty | spovheadcnational | 1230 | 38.52 | 15.13 | 4.1 | 82.3 | 234 | 35.90 | 14.20 | 4.1 | 82.3 | 80.98% |
| Social contributions | ssocialconperofrev | 1230 | 3.23 | 7.53 | 0 | 39.74 | 569 | 3.90 | 8.77 | 0 | 39.74 | 53.74% |

The dataset was constructed in five main steps (also illustrated in Appendix C).

***Step1- Data collection:*** In the first step, I collected 64 variables from publicly available databases (see Appendix B for a complete list of variables and their sources). These variables are potentially crucial for measuring the six capacities of countries. This initial dataset (original) contains a large number of missing values for countries and variables of interest.

***Step2- Choice of Specification:*** To multiply impute, the choice of a correct multiple imputation specification is necessary. In *STATA*, either one can employ multivariate normal (MVN) MI or MI by chained equations (MICE).[5] Both these strategies assume a MAR pattern in data before execution. I argue LMICs exhibit MAR pattern. The pattern, by definition, implies that the observed data can explain and predict missingness (Afghari et al., 2019). LMICs can have missing data for a variety of reasons, ranging from poor data infrastructures and meager resources to frequent natural disasters and severe civil conflicts. However, despite missingness in many variables of significance, LMICs offer rich information on poverty indicators, economic development, literacy rates, and demographics. I argue that this rich corpus of data can be employed to explain and predict the missingness pattern for data on other variables, thus justifying the MAR assumption.

Furthermore, since all the variables are continuous, differently distributed, and missingness among them is arbitrary, Rubin's (1987) multiple imputation by chained equations (MICE) best serves this study. Researchers argue that MICE allows sound modeling for missing values and provides rigorous standard errors for the fitted parameters (Zhang 2016b; White, Royston, and Wood 2011). MICE treats each variable with missing values as the dependent variable in a regression, with the remaining variables as its predictors. Once MICE is specified, as mentioned earlier, within MICE, one can use either a linear regression *(regress)* or predictive mean matching (*PMM*) specification for continuous variables. Chained imputation with *linear regression* has a severe pitfall as it implements normal distribution on imputed values regardless of the distribution of original values (White, Royston, and Wood 2011). Conversely, PMM caters to this problem by respecting the observed values' distribution pattern. Besides, the use of PMM is robust against other misspecifications in the imputation model (Lee and Carlin 2017). Notably, it is robust against heteroskedastic residuals and nonlinear associations between variables (Lee and Carlin 2017; Kleinke 2017). Since the observed variables are not normally distributed (see kernel density graphs plotted after imputation in Appendix) and their residuals are heteroscedastic, PMM is the most suitable chained imputation for this data.

---

[5] One can employ Amelia II in R statistical tool (Honaker, King, and Blackwell 2011). However, Amelia II assumes normality, which is not the case here.

***Step3- Variable shortlisting and running the first round of imputations:*** In the third step, I ran MICE in *STATA 16* for all variables. Out of 64 variables, chained imputations did not work for three variables (multipoverty index, multipoverty intensity, agricultural machinery).[6] I tried linear regression specification too, but it did not work. I dropped them off. Then I run a first successful round of imputations (m=20) followed by descriptive analyses of all these 61 variables. Out of these variables, I dropped off another 14 variables because the results were not of sufficient reliability. They had a considerable fraction of missing information (FMI),[7] or their descriptive statistics were very different from the observed (incomplete) dataset, and they varied a lot in successful imputations. Thus, the list of variables was reduced to 47.

***Step 4- Running the second round of imputations on shortlisted variables:*** In the fourth step, I did a second round of PMM imputations for the truncated list of 47 variables together. I included data on complete variables of time and country identifiers (year and country) and auxiliary variables (GDP per capita, technical cooperation grant, total population, gross capital formation, net ODA and official aid assistance, number of international tourist arrivals receipts, merchandise import from high-income economies as percentage of total merchandize imports, current health expenditure) following the recommendations of the multiple imputation literature.  The inclusion of complete identifiers and other auxiliary variables increases the precision of the imputation results for variables exhibiting high missingness and makes the MAR assumption more plausible (Hardt, Herke, and Leonhart 2012). To obtain a high-efficiency level in parameter results, I set m = 50, i.e., fifty complete datasets (copies of original dataset) were estimated for all 47 variables.[8]

---

[6] The system gave the error message that "the posterior distribution from which MI drew the imputations for these variables is not proper when the VCE estimated from the observed data is not positive definite." This essentially means that there is collinearity. Since these variables have more than 97% missing values, therefore, to deal with the reported error I dropped off these variables from the analysis.

[7] Generally, these variables reported FMI higher than 60%. FMI is the proportion of the total sampling variance that is due to missing data. It is calculated based on the percentage missing for a specific variable and how correlated this variable is with other variables in the imputation model (Pan and Wei 2018). A high FMI shows a problematic variable.

[8] Traditionally researchers set $m = 5$ or 10. New research indicates that $m$ should be high to achieve accurate standard errors and point estimates (von Hippel 2020). With large $m$, variance estimates stabilize, and standard errors become more accurate. In essence, by returning accurate standard errors, large m models the uncertainty within imputations

Subsequent econometric analyses are performed separately on each dataset (50 analyses because m= 50). Then, the results from each analysis are pooled according to Rubin's rules. Here, I randomly pick results from imputation # 25 for descriptive statistics and illustration purposes. This dataset contains 47 variables for 1,230 observations (82 countries for the period 2005-2019).

***Step 5- Quality check:*** Finally, I thoroughly investigated the variables to analyze the imputed values' quality. This investigation informs the extent to which the new complete dataset may be regarded as reliable. I did a visual inspection of kernel density graphs of imputed values, completed values, and original values for all the variables in this investigation. Similarly, I checked descriptive statistics of observed and imputed values. This quality check is discussed fully in the next section. This check results suggest that multiple imputations with PMM have been successful for the truncated list of variables.

In brief, following the above steps, the final version of the MSK database is constructed and made available. The dataset consists of 47 variables for 82 IDA-eligible countries spanning over 15 years (1,230 country-years observations). In contrast, the remaining 17 variables were rejected and not included in the database because either the system could not impute them or returned unreliable imputed values of poor quality.

## 6. Descriptive analysis of the MSK dataset

To empirically illustrate the usefulness of the MSK dataset and how it can be used to study absorptive capacity systems across countries, I have conducted a detailed analysis in another article (Khan 2021). A brief descriptive analysis of the MSK dataset is conducted here. This analysis offers insights into the trends in capacities constituting NACS in LMICs and how they evolve over time. Three brief analyses are conducted: distribution (kernel density) of select few variables of interest within each capacity at the start, middle, and the end of the study period (i.e., 2005, 2010, and 2019); time trends (2005-2019) of the variables of interest for select countries (six countries,

---

(missing values are uncertain) with more certainty. In addition, large *m* is particularly recommended if FMI is high for variables. Similarly, large *m* increases the relative efficiency of parameters (point estimates). i.e., how well the true population parameters are estimated. Generally, when the amount of missing information is high, more imputations (high *m*) are needed to attain adequate efficiency for point estimates (von Hippel 2020; Pan and Wei 2018).

one from each region in our countries of study); and comparative ranking of countries based on composite capacity indices.

### i)   *Distribution (kernel density) of select few variables of interest within each capacity at different periods (i.e., 2005, 2010, and 2019):*

The distribution patterns (Appendix D) are drawn for a select set of variables from each capacity for three years (2005, 2010, and 2019). Distributions for technological capacity by and large show that LMICs have not significantly improved their technological base. A rightward shift in distributions for infrastructure capacity indicates that LMICs overall have experienced an improvement in their infrastructure base. However, we see a leftward shift in the distributions for social capacity, meaning that LMICs eligible for IDA support are moving backward in their social capacity. For the remaining three capacities (human, financial, and public policy), cross-country distributions' evolution is not very evident. Their pattern depends on the specific variable under discussion. For example, distributions for human capacity show that employment in the service sector has improved over time. On the contrary, expenditure on education has not increased.

### ii)   *Time trends (2005-2019) of the variables of interest for select countries (six countries, one from each region in our countries of study):*

Next, time trends of the select variables from each capacity are observed over time for six countries (Appendix E). The trends for technological capacity variables vary over time for different countries. While most of the trends are either uniform or erratic, scientific articles and ECI scores are rising for Pakistan from 2005 onwards. Similar trends (uniform in some cases and unpredictable in others) are observed for financial capacity variables. Myanmar and Nicaragua lead for domestic credit availability, while other countries have experienced an oscillating trend (increasing and then decreasing). In the case of human capacity and infrastructure capacity, trends for some variables (primary completion, expenditure on education, LPI score) have experienced erratic movements; however, most countries are improving in other variables (service and technological sector employment, mobile and internet penetration) of these capacities. This may allude to the fact that these countries are perhaps catching up with advanced economies in terms of these indicators. Finally, it is hard to identify a clear winner for the last two capacities (public policy and social capacity); most trends are either uniform or erratic. However, the statistical score index is strikingly improving for Djibouti and Myanmar. These results largely corroborate the abovementioned distribution analysis. The crux is that countries show varying progress (clearly visible in some cases and diffused in other cases) over time for all these variables.

*iii)*      ***Comparative ranking of countries:***

Lastly, comparative ranking of countries was conducted for recent data in 2019 (see Appendix F). For this, I first calculated six composite indices (Technology, Finance, Human Capital, Infrastructure, Public Policy, and Social Capacity) and then aggregated them into a composite Absorptive Capacity Index. Vietnam tops the list of the countries whereas, South Sudan scored the least. This ranking can be conducted for all years, which would show longitudinal changes in absorptive capacity systems of countries.

While not an exhaustive list of the uses of the dataset, these analyses provided a flavor of how this dataset might be used in comparative analyses of National Absorptive Capacity Systems. These analyses can be extended and conducted in a number of ways in future research. This section's purpose was to give a brief demonstration of how one might get started on subsequent empirical analyses.

## 7.   Quality check of the estimated MSK dataset

A quality check is conducted to determine the usefulness and vitality of this dataset.

As mentioned in section 5, I collected 64 variables to measure countries' capacities to construct the database. After carrying out imputations and evaluation, I shortlisted 47 variables to be included in the dataset for an entire range of 1,230 country-year observations (15 years for 82 countries). The remaining 17 variables were rejected either because the system could not impute them (three variables) or the results produced (14 variables) were not of good quality.

In order to assess the imputation procedure and the reliability of the variables included in the MSK dataset, this article conducts a four-way quality check: first descriptive statistics of the two datasets (complete and observed) is conducted; secondly, distributions of completed and observed datasets are observed; thirdly, correlation tables of the observed and complete variables are compared; and fourthly, trends within imputations and convergence pattern are observed.

*i)*        ***Descriptive statistics of two datasets:***

I looked into means, maximum, minimum, and standard deviation for complete and observed datasets. Table 1 reports a comparison of such descriptive statistics for both datasets. First, the table indicates that means (averages) and standard deviations (variability) for all 47 variables are almost identical. Imputing at the mean might reduce variability in some variables, though (as evident in lower standard deviation values). Secondly, we can see that the complete dataset has the same maxima and minima, and the values are meaningful (no negative numbers on researchers, for instance). Moreover, I inspected relative efficiency values for only imputed variables. This glance of relative efficiency values (above 98% for all variables with m=50) suggested highly efficient point estimates. All this shows that the complete dataset's imputed values are roughly the best approximation of the original sources' missing data.

*ii)*       ***Distribution of compete and observed dataset:***

A detailed distribution assessment is conducted for the two datasets. This is accomplished via visual inspection of kernel densities for all 47 variables in the observed (incomplete) and complete (MSK) datasets.

The logic behind comparing the two datasets statistical distributions is to see how best the complete dataset is an extension of the observed dataset. If the two distributions are roughly similar, we can claim the reliability of the imputed values. But, if the two distributions differ, the imputation results may not be reliable.

Visual inspection of kernel densities provides an interesting quality check (See Appendix G). For almost all the variables within capacities, variables' distributions in the MSK dataset are similar to those in the incomplete data in various imputations.[9] Even for those variables that report missingness higher than 80% (R & D, Researchers, Technicians, Account ownership, HCI scale), the approximation level (similarity), while relatively lower, is still very close to the original distributions. This means that the PMM imputation has successfully estimated missing information with high accuracy. Thus, this visual inspection of kernel density distributions grants substantial reliability status to the MSK dataset.

---

[9] I looked into kernel density distributions at different imputations (randomly chosen) for all capacities.

### iii)    *Correlation tables of the original and complete:*

Lastly, pairwise correlation coefficients are calculated and compared in the original dataset and complete dataset.[10] The tables, shown in Appendix H, report such correlation coefficients for each capacity within both datasets. The correlation coefficients for the observed dataset are reported above the pairwise correlations for the complete dataset.

The rationale behind this correlation comparison is if the two correlations are similar, then statistical distributions between the two will likely match. This will indicate the reliability of the imputation results. However, if the two coefficients are not comparable, this would mean unreliability and bias in the imputation results produced through the imputation procedure. The bias and unreliability will subsequently affect post-imputation analysis on the complete dataset.

A close inspection of the correlation tables suggests that correlation coefficients are very similar across the variables in both datasets. Not only the magnitudes of coefficients are roughly similar, but also the signs of the coefficients are maintained in the complete dataset following the multiple imputation exercise. Some coefficients (for example, R & D, Number of technicians, Domestic credit, among others) change in size; however, these changes are not substantial. Overall, this check suggests that PMM imputation has preserved the correlation structure among the variables. Thus, it can be concluded that the MSK dataset is sufficiently reliable.

### iv)    *Trends within imputations and convergence pattern:*

Similarly, I inspected the trends in imputed variables' values across imputations (at m=1, m=10, m=25, m=40, m=50). I noticed that values across imputations were highly similar, suggesting that the imputation exercise was successful. Also, since the dataset was obtained through chained imputations involving iterations, the reliability of the imputation process must be established. Therefore, to establish the reliability of the imputation process, I checked for convergence among iterations for imputed variables. The convergence pattern of the iterations through which the dataset was generated showed a healthy convergence (Appendix I). A healthy convergence means that variance between and within iterations is the same. All this shows that the MSK dataset is of good quality.

---

[10] Correlations were compared of original (m=0) and complete dataset (at imputation m=25).

## 8. Conclusion and Implications

Comparative country analyses on absorptive capacity and economic development in LMICs lack because of the lack of complete data availability. To address this problem, this article employed Rubin's Multiple Imputation to impute missing values in variables. Specifically, it used Multiple Imputation by Chained Equations with Predictive Mean Matching approach to estimate the MSK panel dataset. The dataset consisted of six country-related capacities. A total of 47 continuous variables measured these capacities. This dataset was estimated from an observed dataset containing a lot of missing values. The complete dataset contained 82 countries for the period 2005-2019, for 1,230 country-year observations.

The MSK dataset provides a rich panel (across countries and over time) of statistical content that can be used in several ways. For instance, this dataset can be used to estimate the impact of absorptive capacities on economic growth in LMICs. Similarly, the capacities can be aggregated for different LMICs to find the relative standing of one economy viz-a-viz other economies. Further, such an exercise can be used to investigate the factors of development within leading and lagging LMICs. Finding leading and lagging economies within LMICs at the same level of development offer lessons to lagging economies on how they can catch up. Here, I demonstrated how a simple descriptive analysis of capacities within the complete dataset could be used to gain insights into the dynamic evolution of such capacities in different countries.

On the methodological front, MICE PMM for estimating dataset for the comparative analyses of capacities and economic growth in LMICs is powerful compared to other solutions such as mean imputation or deletion. MICE PMM is powerful because it retains variability in data as the imputed value is randomly taken from the suitable donor pool. Moreover, PMM is a good technique because it reduces bias by keeping information on all variables (variables for which partial data is available are imputed rather than deleted). Similarly, the technique preserves representation (by keeping all economies even if they have partial data rather than dropping them of analysis), returns accurate or realistic data (imputed data is taken from neighboring data pool), and captures dynamic evolution for all economies (which is compromised by using other imputation techniques).

However, the fact that MI returns multiple datasets establishes the uncertainty underlying the values of missing data. No matter how rigorous it is, no imputation can claim with 100 percent certainty the accuracy of imputed values. Therefore, the dataset generated through MI must be carefully used for any analysis. The results of such analysis must make a disclaimer about the process through which the dataset was obtained. The reliability or quality check must be performed on the newly generated dataset, just as I did for the MSK dataset. The MSK dataset generated here passed the quality check as the observed and complete dataset exhibited almost similar distributions, descriptive statistics, and correlation coefficients, and the process through which the dataset was imputed returned a healthy convergence among iterations.

As the MI-generated dataset is reliable, the current consensus seems to be that such a dataset is undoubtedly valuable for hypothesis generation. However, what is uncertain is whether a researcher can use such a dataset to estimate the actual quantitative size of the impact (effect size) of one capacity on economic development? One way would be to compare the results based on the original dataset (albeit biased) with those based on the complete dataset and investigate the trends. This may give some insights into the relative vitality of the completed dataset, but as the saying goes, research is needed.

## Appendix A: Handling Missing Data Strategies, Assumptions, Advantages and Disadvantages

| Strategies | Definition | Assumption | Advantage | Disadvantage |
|---|---|---|---|---|
| **Listwise deletion** | Complete-case analysis). It removes all data for a case that has any missing values (Kang 2013; Donner 1982) | MCAR | -Generally used if the researcher is performing a treatment study and wishes to compare a completers analysis (listwise deletion) vs. an intent-to-treat analysis (includes cases with missing data imputed or considered in a treatment design)<br>- Can be applied to any statistical model (structural equation modeling, multi-level regression, etc.)<br>- In the instance of MAR among independent variables (i.e., they do not depend on the values of dependent variables), listwise deletion parameter estimates can be unbiased. (Little 1992) | - MCAR assumptions are generally rare to support<br>- Produce bias parameters and the estimates |
| **Pairwise deletion** | Available-case analysis aims to reduce the loss that occurs in listwise deletion. Pairwise maximizes all data available through checking into the correlation matrix between variables (Kang 2013; J.-O. Kim and Curry 1977) | MCAR | -It increases statistical power in analyses<br>- Could be used in linear models such as linear regression, factor analysis, or SEM. | - Produce under- or overestimated standard of errors<br>- If the data mechanism is MAR, pairwise will return biased estimates. |
| **Mean substitution** | This method substitutes the mean value of a variable for missing value (Kang 2013; Zhang 2016a). Also called unconditional mean substitution | NA | -Simple to execute | - Does not preserve the relationships among variables<br><br>- Leads to underestimated standard errors |
| **Regression imputation** | Called as conditional mean imputation, here missing value is based (regressed) on other variables (Zhang 2016a) | -MCAR or MAR | -Maintain the relationship with other variables<br>- If the data are MCAR, least-squares coefficients estimates will be consistent and unbiased in large samples (Gourieroux and Monfort 1981) | - No variability left<br>-Treated data as if they were collected<br>- Leads to underestimated standard errors & overestimated test statistics |
| **Cold deck imputation** | Cold Deck picks value from a case that has similar values on other | MAR | -Easy to execute | -Removes the desired random variation |

| Strategies | Definition | Assumption | Advantage | Disadvantage |
|---|---|---|---|---|
| | variables (Haukoos & Newgard 2007) | | | |
| **Maximum likelihood (ML)** | It models the missing data based on observed data. This procedure considers available data as part of some distribution. Subsequently, parameters are estimated that maximize the chance of observing the observed data (Enders 2001) | -MAR and Monotonic (meaning, that if an observation's is missing on one variable, then the following variables of that observation have also missing data | - Consistent<br>-Asymptotically efficient (becomes efficient for large sample)<br>-Asymptotically normal | - ML can usually handle linear models, log-linear models. However, beyond that, ML still is lacking in theory and software implementation |
| **Expectation-Maximization Algorithm** | Similar to ML, but it is an iterative process. In the Expectation stage, data is imputed from observed data. In the second stage, the values are checked if they are the most likely. If not, it imputes again a more likely value (Enders 2001) | MAR | - Easy to use<br>- Preserves the relationship with other variables | - Standard errors of the coefficients are incorrect (biased usually downward - underestimate)<br><br>-Models with overidentification, the estimates will not be efficient |
| **Multiple imputation**<br><br>**(many ways to execute MI)**<br><br>**-Multivariate Normal MI**<br>**-Chained MI  -**<br>**(Predictive Mean Matching, Regression, Logistic)** | MI replaces missing values with a set of imputed values. Analyses are subsequently performed on all the imputed values, and results are pooled (Rubin 1987) | MAR | -Consistent<br><br>-Asymptotically efficient<br>-Asymptotically normal<br>- MI can be applied to any model, unlike ML, which can be applied only to limited models | - MI delivers a little different result in various runs. By seeding, the problem can be evaded.<br>- Some MI methods may cause unlikely values (e.g., negative values)<br>- Not all MI methods can handle heteroskedastic data |

**Appendix B: List of all 64 Variables, their Definitions, Sources, Missingness Amount in Observed Variables, and Acceptance/Rejection Status for the MSK Dataset**

| Capacity | Variable code | Definition and source of the variables included in the MSK Database | | | |
|---|---|---|---|---|---|
| **Capacity** | **Variable code** | **Definition** | **Source** | **%Missing** | **Accept/Reject** |
| **Technology Capacity** | tippay | **Charges for the use of intellectual property, payments (BoP, current US$).** Payment or charges per authorized use of intangible, non-produced, non-financial assets and proprietary rights (such as patents, trademarks, copyrights, industrial processes and designs including trade secrets, and franchises) and for the use, through licensing agreements, of produced originals of prototypes. Data are in current US dollars | IMF, World Bank | 33.50% | Accepted |
| | tinddesapprebyco | **Industrial design applications, resident, by count** | WIPO | 75.12% | Rejected |
| | tscitjar | **Scientific and technical journal articles.** Number of scientific and engineering articles published in the following fields: physics, biology, chemistry, mathematics, clinical medicine, biomedical research, engineering and technology, and earth and space sciences, per million people. | World Bank | 6.67% | Accepted |
| | trandd | **Research and development expenditure (% of GDP)** | UNESCO | 81.71% | Accepted |
| | tresinrandd | **Researchers in R&D (per million people)** | UNESCO | 87.97% | Accepted |
| | ttechinrandd | **Technicians in R&D (per million people)** | UNESCO | 88.29% | Accepted |
| | tpatappre | **Patent applications, residents** | WIPO | 60% | Rejected |
| | ttradappresbyco | **Trademark applications, resident, by count** | WIPO | 70.89% | Rejected |
| | thigexperofmanex | **High-technology exports (% of manufactured exports).** High-technology exports are products with high R&D intensity, such as in aerospace, computers, pharmaceuticals, scientific instruments, and electrical machinery. | UN, COMTRADE | 55.53% | Accepted |
| | tsecedvoc | **Secondary education, vocational pupils.** Secondary students enrolled in technical and vocational education programs, including teacher training. | UNESCO | 53.58% | Accepted |
| | teciscore | **ECI Score.** Measure of economic complexity containing information about both the diversity of a country's export and their sophistication. High ECI Score shows that an economy exports many goods that are of low ubiquity and that are produced by highly diversified countries. In other words, diverse and sophisticated economies have high scores. | OEC, MIT | 27.48% | Accepted |
| **Capacity** | **Variable code** | **Definition** | **Source** | **%Missing** | **Accept/Reject** |
| **Financial Capacity** | fdaystoenfctt | **Time required to enforce a contract (days).** Days required to enforce a contract, whereas the days are counted from the day a plaintiff files the lawsuit in court until payment. Low values indicate high competitiveness and vice verca. | World Bank, Doing Business Project | 6.18% | Accepted |
| | fdomcrprsecbybkpergdp | **Domestic Credit by Banking Sector.** This includes all credit to various sectors (monetary authorities, banks, financial corporations) on a gross basis, with the exception of credit to the central government, which is net, as a % of GDP. | IMF, World Bank | 10.57% | Accepted |
| | fopenind | **Openness Indicator.** (Import + Export)/GDP. Constant US 2010. | World Bank | 31.14% | Accepted |
| | fdepcombkp1k | **Depositors with commercial banks (per 1,000 adults)** | IMF, World Bank | 47.32% | Rejected |
| | fdaystoregpro | **Time required to register property (days).** The number of calendar days needed for businesses to secure rights to property. | World Bank, Doing Business Project | 10.24% | Accepted |
| | fcosbstpropergni | **Cost of business start-up procedures (% of GNI per capita)** | World Bank | 6.18% | Accepted |
| | ftaxrpergdp | **Tax revenue (% of GDP).** Tax revenue means compulsory transfers to the government for public purposes. | IMF, World Bank | 52.60% | Accepted |
| | fcombkbr1k | **Commercial bank branches (per 100,000 adults)** | IMF, World Bank | 10.65% | Accepted |
| | fdaystoobteleconn | **Time to obtain electrical connection (Days).** Days to obtain electrical connection. Days experienced to obtain an electrical connection from the day an establishment applies for it to the day it receives the service. | World Bank, Enterprise Survey | 87.56% | Accepted |
| | ftdaystobusi | **Time required to start a business (Days).** The number of days needed to complete the procedures to legally operate a business. | World Bank, Doing Business Project | 6.18% | Accepted |
| | faccownperofpop15p | **Account ownership at a financial institution or with a mobile-money-service provider (% of pop ages 15+).** Account denotes the percentage of respondents who report having an account (by themselves or together with someone else) at a bank or another type of financial institution or report personally using a mobile money service in the past 12 months (% age 15+). | Demirguc-Kunt et al., 2018, Global Financial Inclusion Database, World Bank. | 86.99% | Accepted |
| | fnewbusdenper1k | **New business density (new registrations per 1,000 people ages 15-64).** New businesses registered are the number of new limited liability corporations registered in the calendar year. | World Bank, Enterprise Survey | 52.60% | Accepted |

| Capacity | Variable code | Definition | Source | %Missing | Accept/Reject |
|---|---|---|---|---|---|
| **Human Capacity** | hprimenrollpergross | **School enrollment, primary (% gross).** Ratio of total enrollment, regardless of age, to the population of the age group that officially corresponds to the primary level. | UNESCO | 25.93% | Accepted |
| | hsecenrollpergross | **School enrollment, secondary (% gross).** Ratio of total enrollment, regardless of age, to the population of the age group that officially corresponds to the secondary level. | UNESCO | 42.20% | Accepted |
| | hcompeduyears | **Compulsory education, duration (years).** No. of years that children are legally obliged to attend school. | UNESCO | 16.42% | Accepted |
| | hgvtexpedupergdp | **Government expenditure on education (% of GDP).** General government expenditure on education (current, capital, and transfers) is expressed as a percentage of GDP. | UNESCO | 50% | Accepted |
| | hpupteapriratio | **Primary pupil-teacher ratio.** Ratio (number of pupils enrolled in primary school) / (number of primary school teachers) | UNESCO | 38.94% | Accepted |
| | hempinduspertotem | **Employment in industry (% of total employment).** Employment is defined as persons of working age who were engaged in any activity to produce goods or provide services for pay or profit, whether at work during the reference period or not at work due to temporary absence from a job, or to working-time arrangement. The industry sector consists of mining and quarrying, manufacturing, construction, and public utilities (electricity, gas, and water), in accordance with divisions 2-5 (ISIC 2) or categories C-F (ISIC 3) or categories B-F (ISIC 4). | ILO, World Bank | 8.54% | Accepted |
| | hempserpertotem | **Employment in services (% of total employment).** Employment is defined as persons of working age who were engaged in any activity to produce goods or provide services for pay or profit, whether at work during the reference period or not at work due to temporary absence from a job, or to working-time arrangement. The services sector consists of wholesale and retail trade and restaurants and hotels; transport, storage, and communications; financing, insurance, real estate, and business services; and community, social, and personal services, in accordance with divisions 6-9 (ISIC 2) or categories G-Q (ISIC 3) or categories G-U (ISIC 4). | ILO, World Bank | 8.54% | Accepted |
| | hprimcompra | **Primary completion rate, total (% of relevant age group)** | UNESCO | 40.24% | Accepted |
| | hhciscale0to1 | **Human capital index (HCI) (scale 0-1).** The HCI calculates the contributions of health and education to worker productivity. The final index score ranges from zero to one and measures the productivity as a future worker of child born today relative to the benchmark of full health and complete education. | World Bank | 87.48% | Accepted |
| | hlfwithadedu | **Labor force with advanced education (% of total working-age population with advanced education)** | ILO, World Bank | 78.46% | Accepted |
| | hlfwithbasiced | **Labor force with basic education (% of total working-age population with basic education)** | ILO, World Bank | 78.13% | Rejected |
| | hlfwithintermeded | **Labor force with intermediate education (% of total working-age population with basic education)** | ILO, World Bank | 78.13% | Rejected |

| Capacity | Variable code | Definition | Source | %Missing | Accept/Reject |
|---|---|---|---|---|---|
| **Infrastructural Capacity** | ielecconkwhpercapita | **Electric power consumption (kWh per capita).** Production of power plants and combined heat and power plants less transmission, distribution, and transformation losses and own use by heat and power plants. | IEA, World Bank | 66.42% | Rejected |
| | icarrierdepwdwide | **Air transport, registered carrier departures worldwide.** Registered carrier departures worldwide are domestic takeoffs and takeoffs abroad of air carriers registered in the country | World Bank | 37.40% | Rejected |
| | imobsubper100 | **Mobile cellular subscriptions (per 100 people).** | International Telecom Union, World Bank | 0.89% | Accepted |
| | itelesubper100 | **Fixed telephone subscriptions (per 100 people)** | International Telecom Union, World Bank | 0.98% | Accepted |
| | ibdbandsubper100 | **Fixed broadband subscriptions (per 100 people)** | International Telecom Union, World Bank | 9.43% | Accepted |
| | iaccesselecperpop | **Access to electricity (% of population).** The percentage of population with access to electricity. | World Bank, Sustainable Energy for All | 7.72% | Accepted |
| | ienergyusepercap | **Energy use (kg of oil equivalent per capita).** The use of primary energy before transformation to other end-use fuels, which is equal to indigenous production plus imports and stock changes, minus exports and fuels supplied to ships and aircraft engaged in international transport. | IEA, World Bank | 61.71% | Accepted |
| | ieletandislossesperoutput | **Electric power transmission and distribution losses (% of output)** | IEA, World Bank | 67.32% | Rejected |
| | imachtpeqpervaladdmanu | **Machinery and transport equipment (% of value added in manufacturing).** Value added in manufacturing is the sum of gross output less the value of intermediate inputs used in production for industries classified in ISIC major division D. Machinery and transport equipment correspond to ISIC divisions 29, 30, 32, 34, and 35. | UNIDO, World Bank | 75.28% | Rejected |
| | iindintperpop | **Individuals using the internet (% of population).** Internet users are individuals who have used the Internet (from any location) in the last 3 months. The Internet can be used via a computer, mobile phone, personal digital assistant, games machine, digital TV etc. | International Telecom Union, World Bank | 1.71% | Accepted |
| | iraillinestotalkm | **Rail lines (total route km).** Railway route in km for train service, irrespective of the number of parallel tracks. | International Union of Railway | 78.05% | Rejected |
| | isecinterserper1mill | **Secure internet servers per 1 million people** | World Bank | 35.93% | Rejected |
| | iagmachtracper100sqkm | **Agricultural machinery, tractors per 100 sq. km of arable land** | FAO, World Bank | 98.05% | Rejected |
| | ilpiquoftratraninfr | **Logistics performance index: Quality of trade and transport-related infrastructure (1=low to 5=high).** Logistics professionals' perception of country's quality of trade and transport related infrastructure (e.g. ports, railroads, roads, information technology), on a rating ranging from 1 (very low) to 5 (very high). Scores are averaged across all respondents. | World Bank | 69.76% | Accepted |
| | | | | | |
| **Capacity** | **Variable code** | **Definition** | **Source** | **%Missing** | **Accept/Reject** |
| **Public Policy Capacity** | pcpiapsmgandinscl1to6 | **CPIA public sector management and institutions cluster average (1=low to 6=high).** The public sector management and institutions cluster includes property rights and rule-based governance, quality of budgetary and financial management, efficiency of revenue mobilization, quality of public administration, and transparency, accountability, and corruption in the public sector. | World Bank, CPIA Database | 7.97% | Accepted |

| Capacity | Variable code | Definition | Source | %Missing | Accept/Reject |
|---|---|---|---|---|---|
| | pcpiastpolclavg1to6 | **CPIA structural policies cluster average (1=low to 6=high).** The structural policies cluster includes trade, financial sector, and business regulatory environment | World Bank, CPIA Database | 7.97% | Accepted |
| | pstrengthoflegalright | **Strength of legal rights index (0=weak to 12=strong).** Strength of legal rights index measures the degree to which collateral and bankruptcy laws protect the rights of borrowers and lenders and thus facilitate lending. The index ranges from 0 to 12, with higher scores indicating that these laws are better designed to expand access to credit. | World Bank, Doing Buisness Project | 54.07% | Accepted |
| | iscapscoravg | **Overall level of statistical capacity (scale 0 - 100).** A composite score (on a scale of 0-100) which assesses the capacity of a country's statistical system in three areas (25 criteria): methodology; data sources; and periodicity and timeliness. | World Bank | 1.95% | Accepted |
| | pcpiaeconmgtcl1to6 | **CPIA economic management cluster average (1=low to 6=high).** The economic management cluster includes macroeconomic management, fiscal policy, and debt policy. | World Bank, CPIA Database | 7.97% | Accepted |
| | | | | | |
| **Capacity** | **Variable code** | **Definition** | **Source** | **%Missing** | **Accept/Reject** |
| | scpiabdhumanres1to6 | **CPIA building human resources rating (1=low to 6=high).** Building human resources assesses the national policies and public and private sector service delivery that affect the access to and quality of health and education services, including prevention and treatment of HIV/AIDS, tuberculosis, and malaria. | World Bank, CPIA Database | 7.97% | Accepted |
| | scpiaeqofpbresuse1to6 | **CPIA equity of public resource use rating (1=low to 6=high).** Equity of public resource use assesses the extent to which the pattern of public expenditures and revenue collection affects the poor and is consistent with national poverty reduction priorities | World Bank, CPIA Database | 7.97% | Accepted |
| | scpiasocprorat1to6 | **CPIA social protection rating (1=low to 6=high).** Social protection and labor assess government policies in social protection and labor market regulations that reduce the risk of becoming poor, assist those who are poor to better manage further risks, and ensure a minimal level of welfare to all people. | World Bank, CPIA Database | 8.29% | Accepted |
| | scpiapolsocincl1to6 | **CPIA policies for social inclusion/equity cluster average (1=low to 6=high).** The policies for social inclusion and equity cluster includes gender equality, equity of public resource use, building human resources, social protection and labor, and policies and institutions for environmental sustainability | World Bank, CPIA Database | 8.29% | Accepted |
| | scovofsocprolbrpro | **Coverage of social protection and labor programs (% of population).** Coverage of social protection and labor programs (SPL) shows the percentage of population participating in social insurance, social safety net, and unemployment benefits and active labor market programs | World Bank | 87.48% | Rejected |
| | sginiinedxwbest | **GINI index (World Bank estimate).** Measures income inequality. A Gini index of 0 represents perfect equality, while an index of 100 implies perfect inequality. | World Bank | 80.16% | Rejected |
| | spovheadcnational | **Poverty headcount ratio at national poverty lines (% of population).** National poverty headcount ratio is the percentage of the population living below the national poverty line(s) | World Bank | 80.98% | Accepted |
| | smultipovertyintensity | **The average share of weighted deprivations (intensity).** | World Bank | 97.97% | Rejected |
| | ssocialconperofrev | **Social contributions (% of revenue).** Social contributions include social security contributions by employees, employers, and self-employed individuals, and other contributions whose source cannot be determined. They also include actual or imputed contributions to social insurance schemes operated by governments | IMF, World Bank | 53.74% | Accepted |
| | smultipoverindex | **Multidimensional poverty index (scale 0-1).** Proportion of the population that is multidimensionally poor adjusted by the intensity of the deprivations | World Bank | 98.78% | Rejected |
| | | | | | |

(The **Social Capacity** label spans the lower block of rows from scpiabdhumanres1to6 through smultipoverindex.)

**Appendix C: Construction of the MSK Dataset**

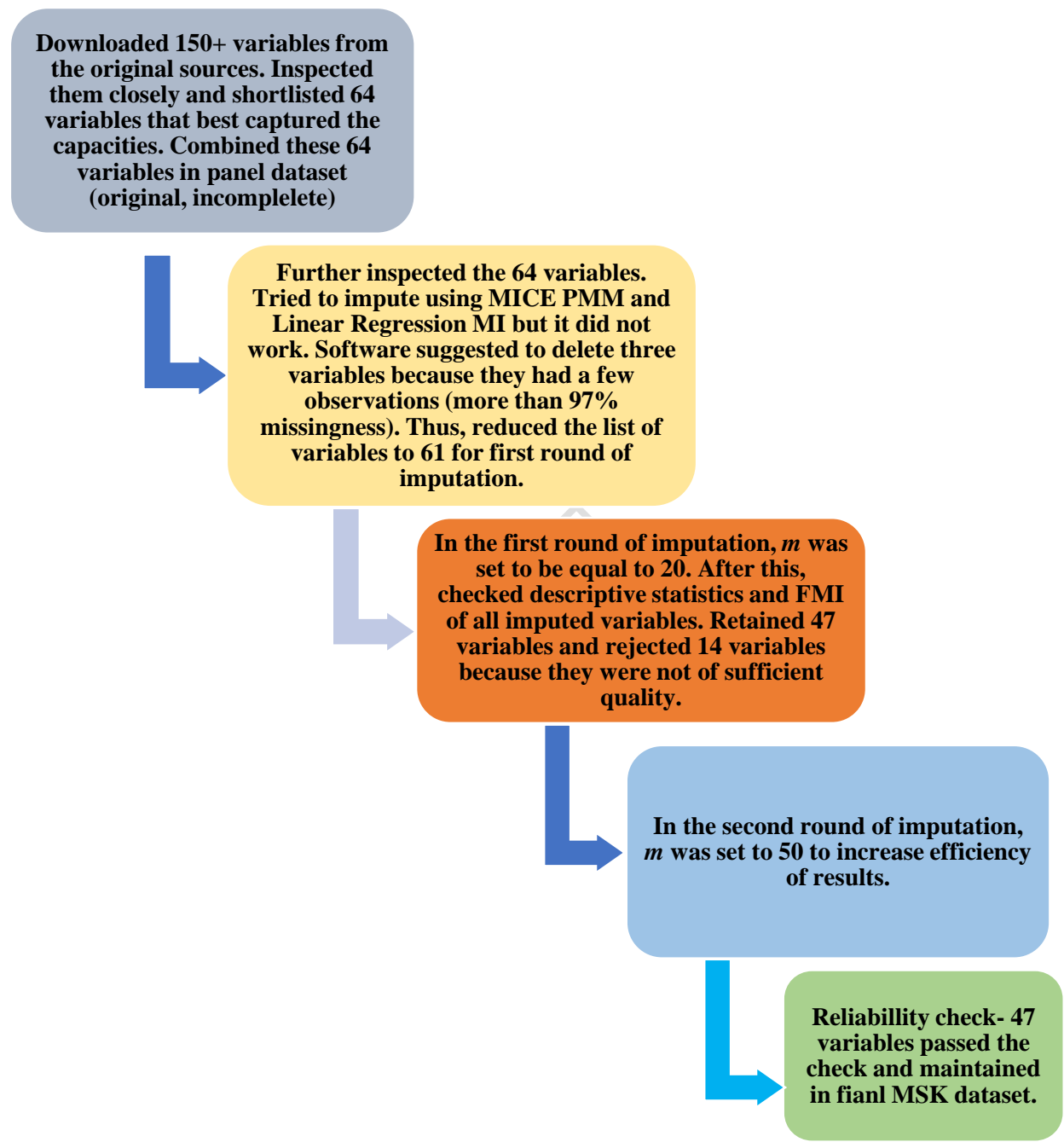

Downloaded 150+ variables from the original sources. Inspected them closely and shortlisted 64 variables that best captured the capacities. Combined these 64 variables in panel dataset (original, incomplelete)

Further inspected the 64 variables. Tried to impute using MICE PMM and Linear Regression MI but it did not work. Software suggested to delete three variables because they had a few observations (more than 97% missingness). Thus, reduced the list of variables to 61 for first round of imputation.

In the first round of imputation, *m* was set to be equal to 20. After this, checked descriptive statistics and FMI of all imputed variables. Retained 47 variables and rejected 14 variables because they were not of sufficient quality.

In the second round of imputation, *m* was set to 50 to increase efficiency of results.

Reliabillity check- 47 variables passed the check and maintained in fianl MSK dataset.

**Appendix D: Kernel Densities for Select variables of Interest at Different Points.**

**Figure D.1. Technology Capacity:**

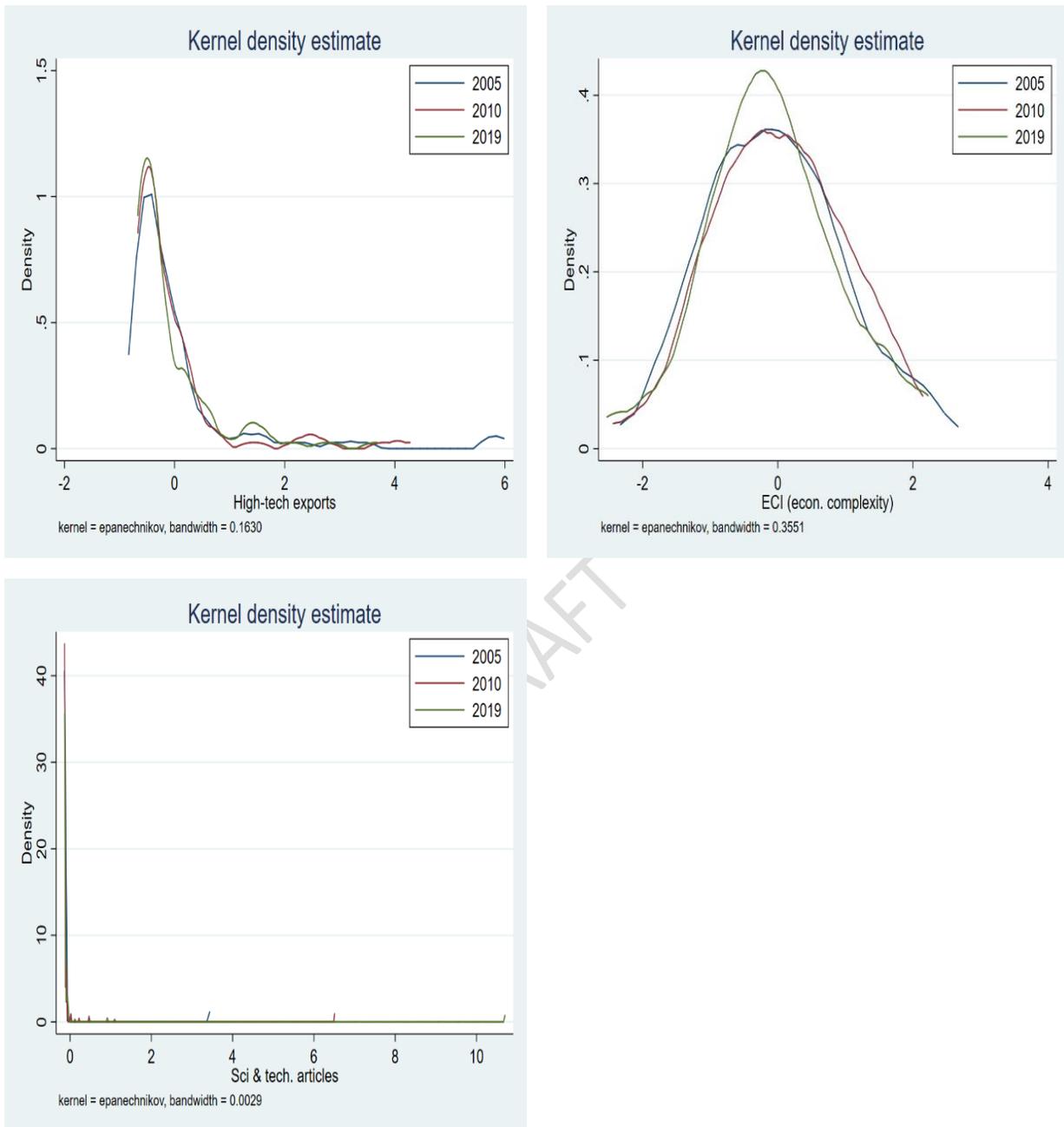

**Figure D.2: Financial Capacity**

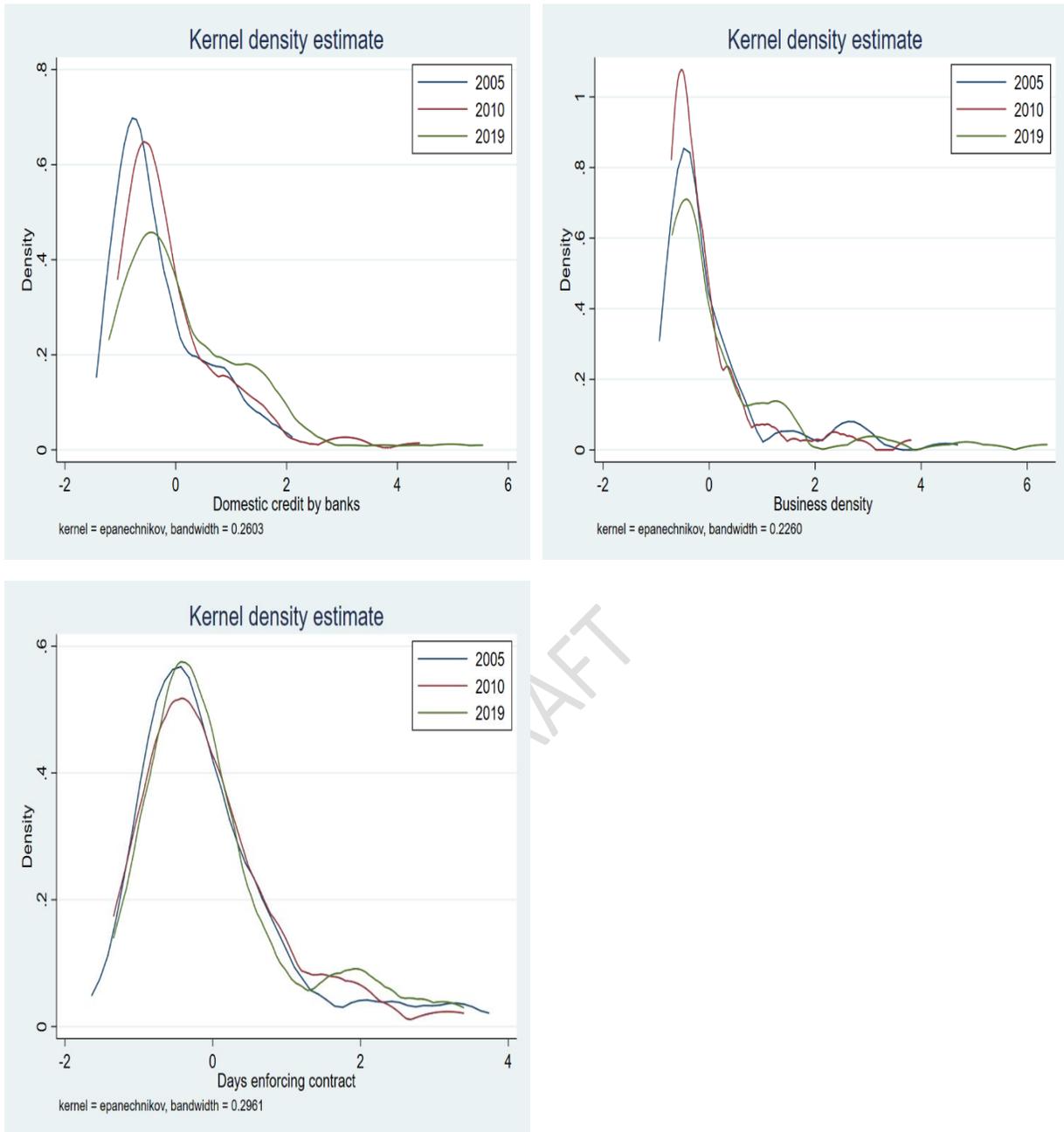

**Figure D.3: Human Capacity**

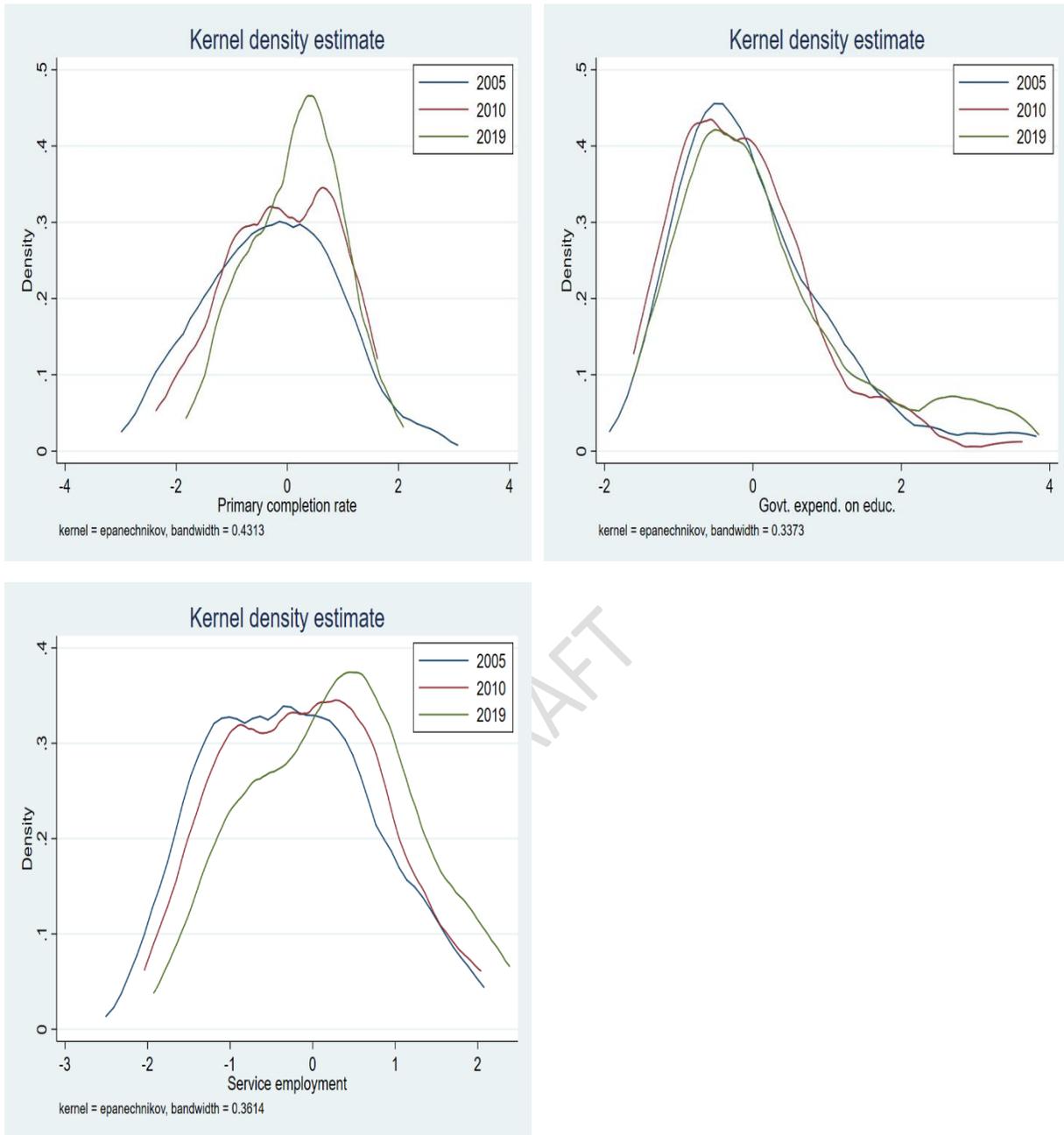

**Figure D.4: Infrastructure Capacity**

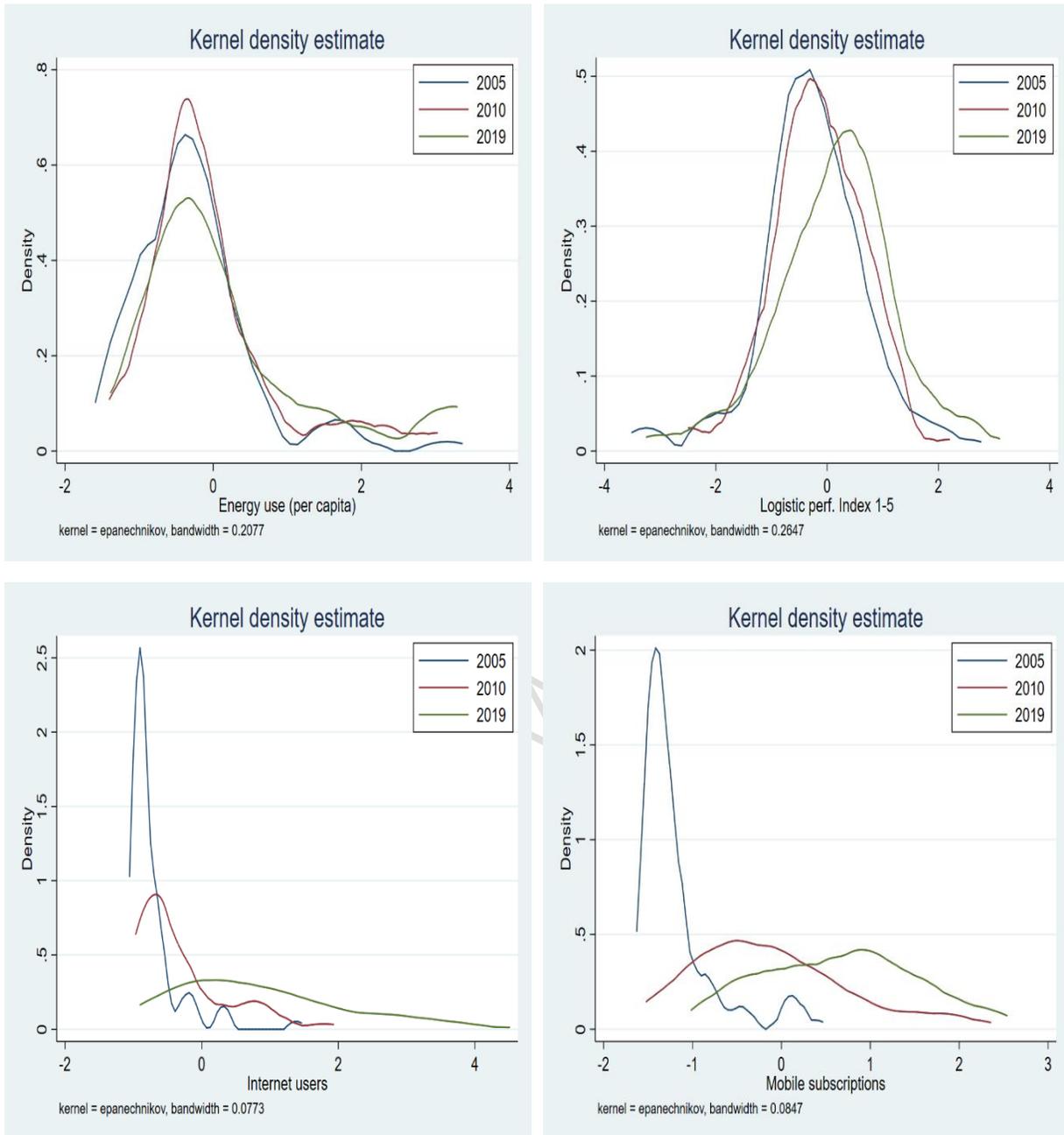

**Figure D.5: Public Policy Capacity**

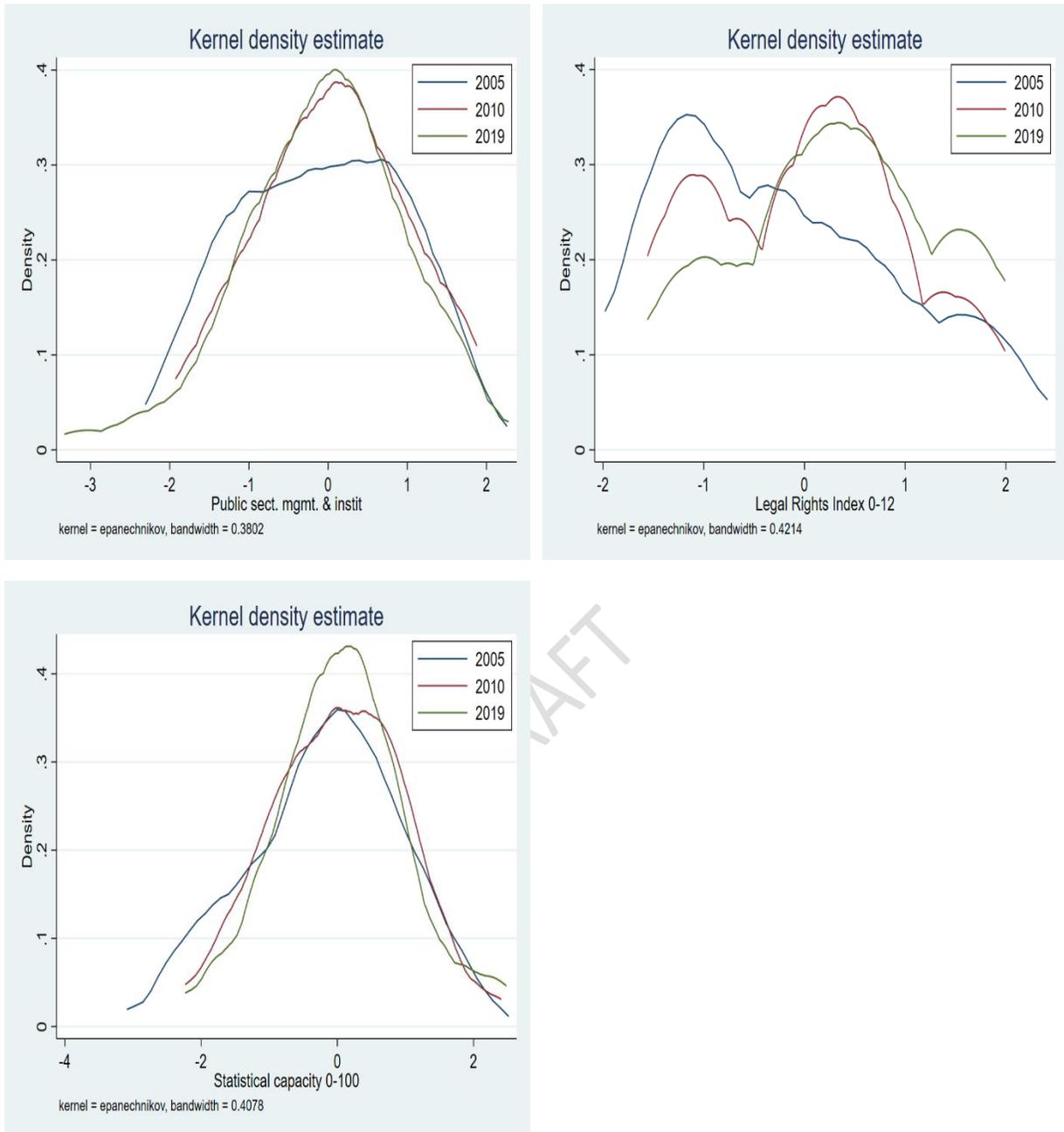

**Figure D.6: Social Capacity**

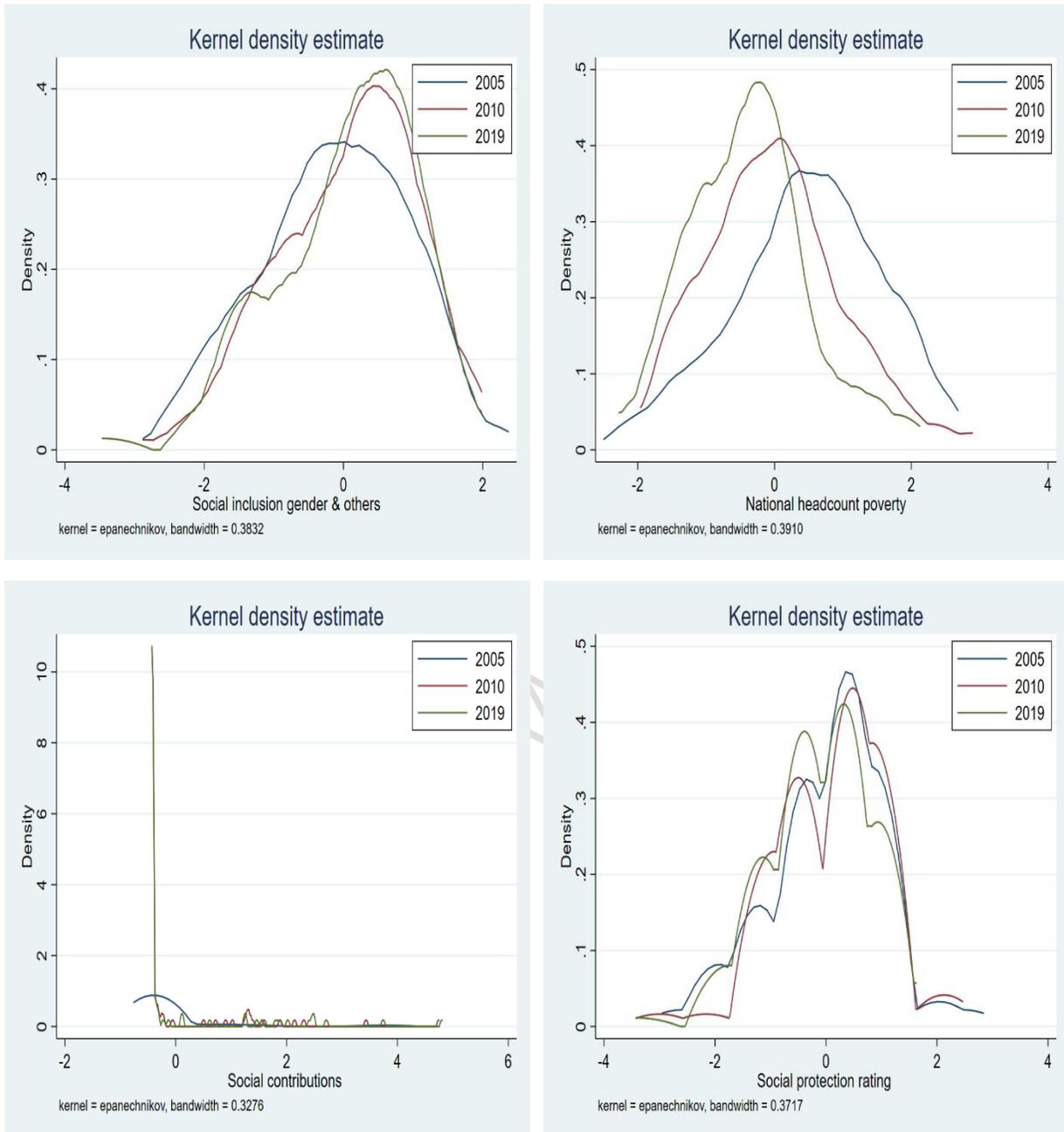

**Appendix E: Time Trends for Select Countries for Select Variables.**

**Figures E.1: Technology Capacity**

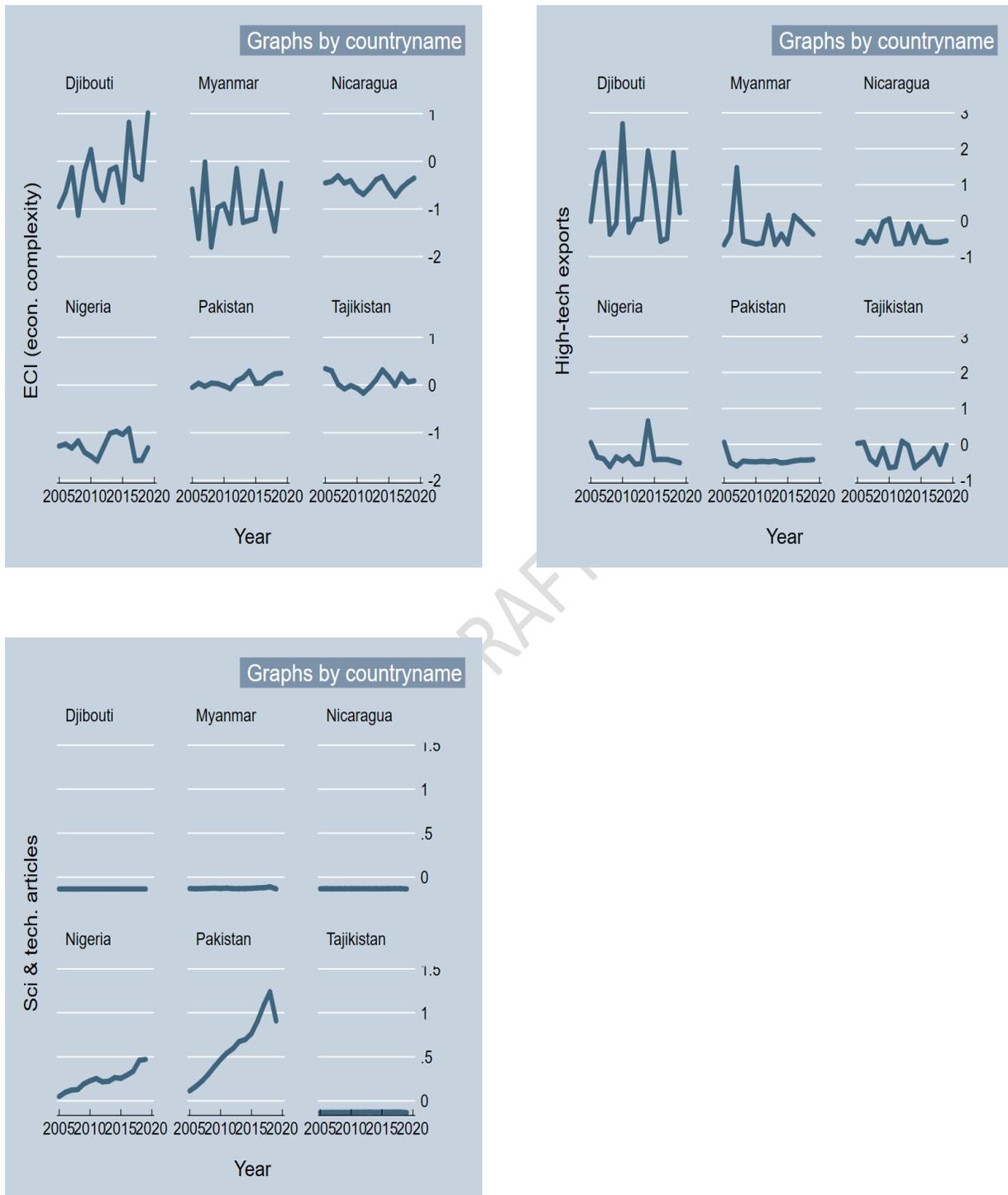

**Figures E.2: Financial Capacity**

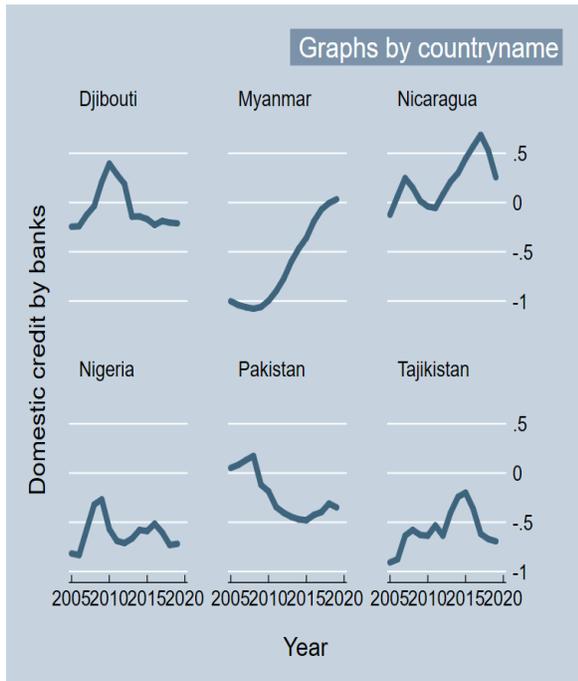

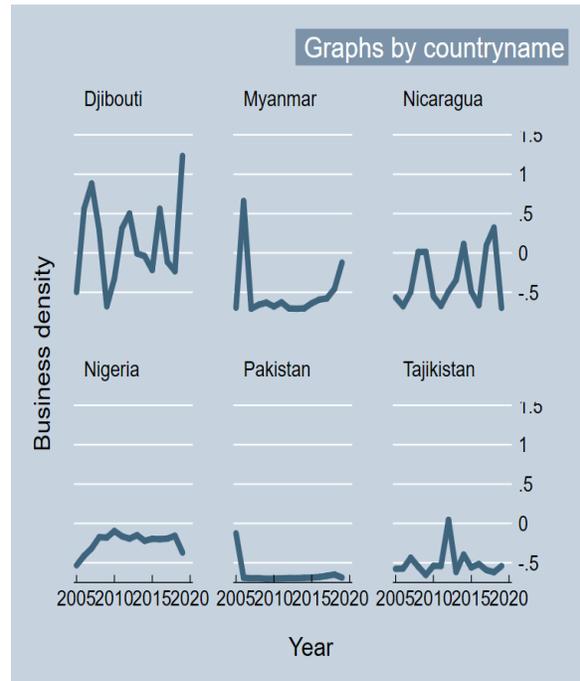

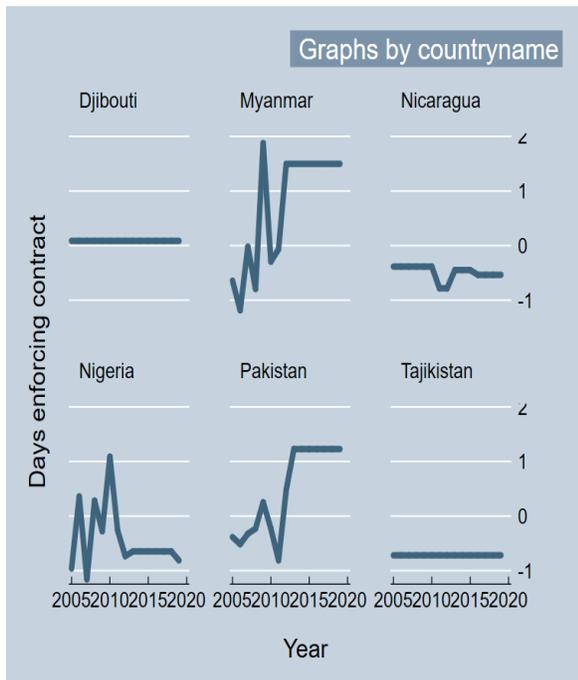

**Figures E.3: Human Capacity**

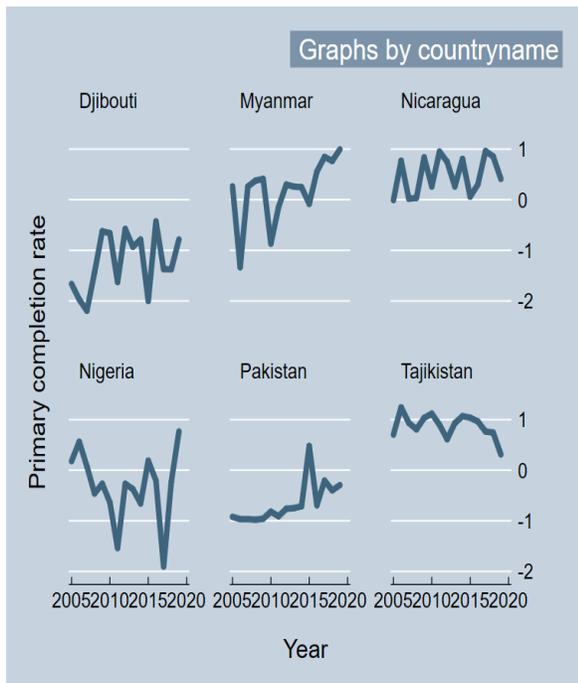

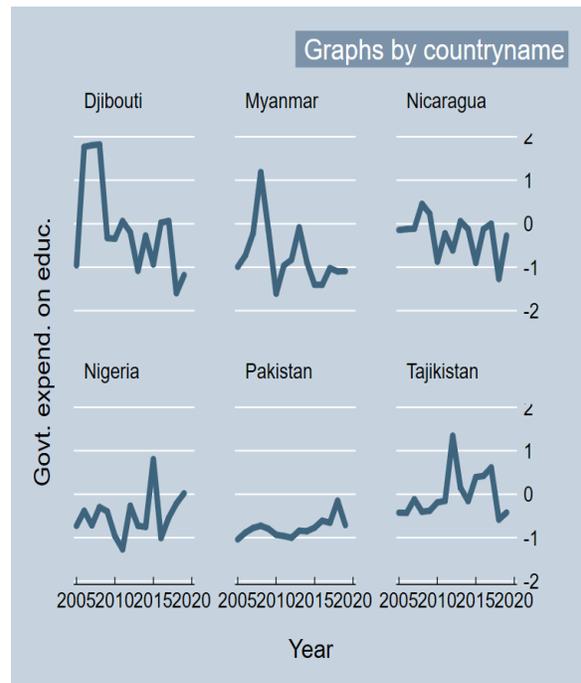

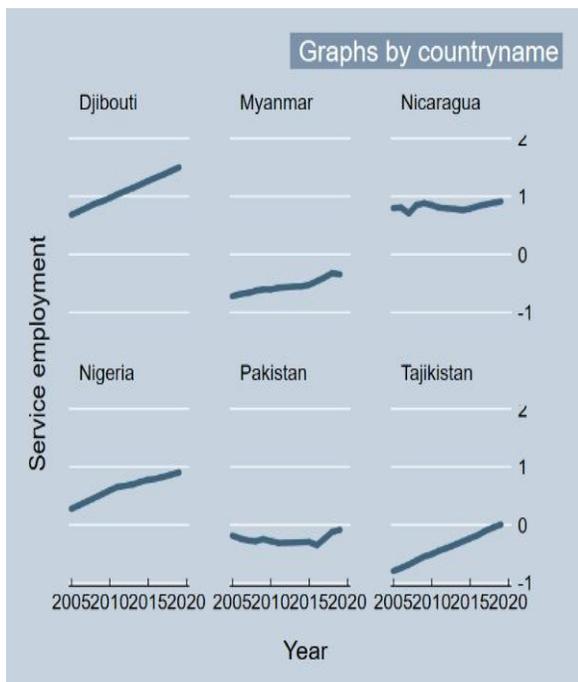

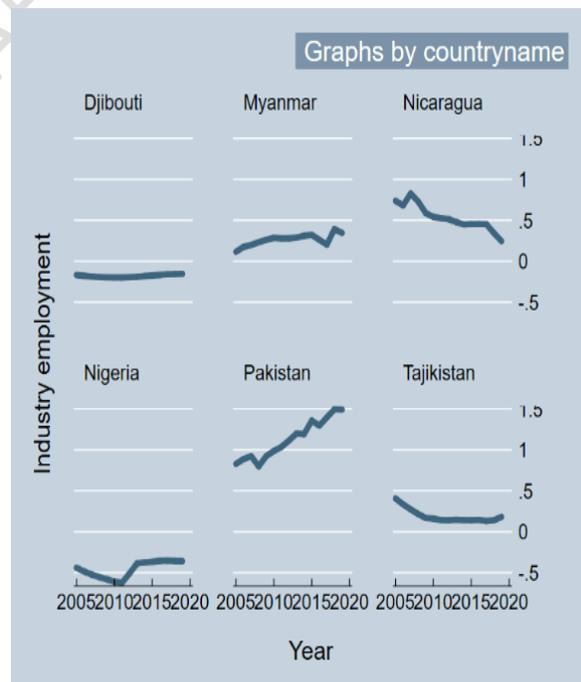

**Figure E.4: Infrastructure Capacity**

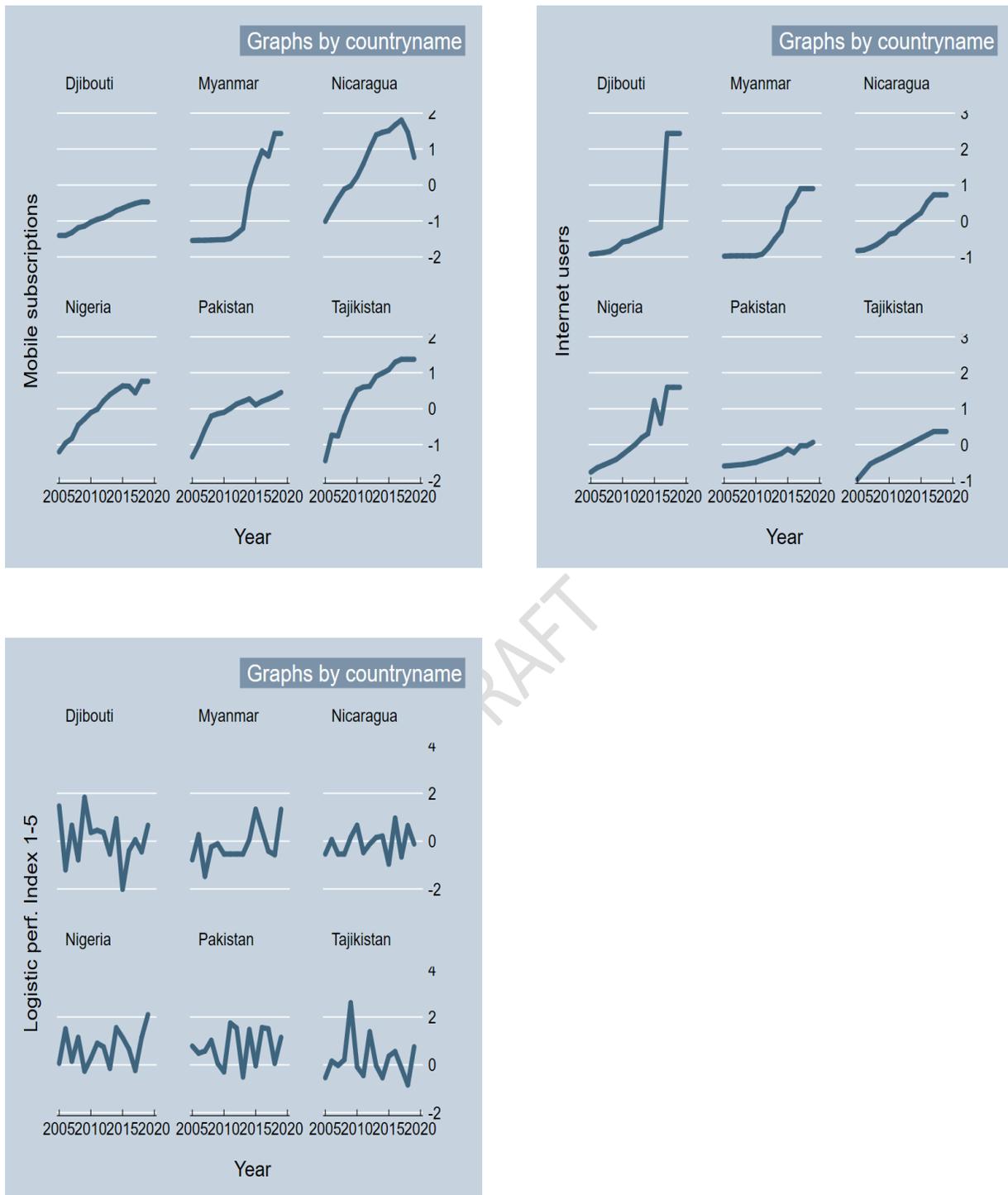

**Figure E.5: Public Policy Capacity**

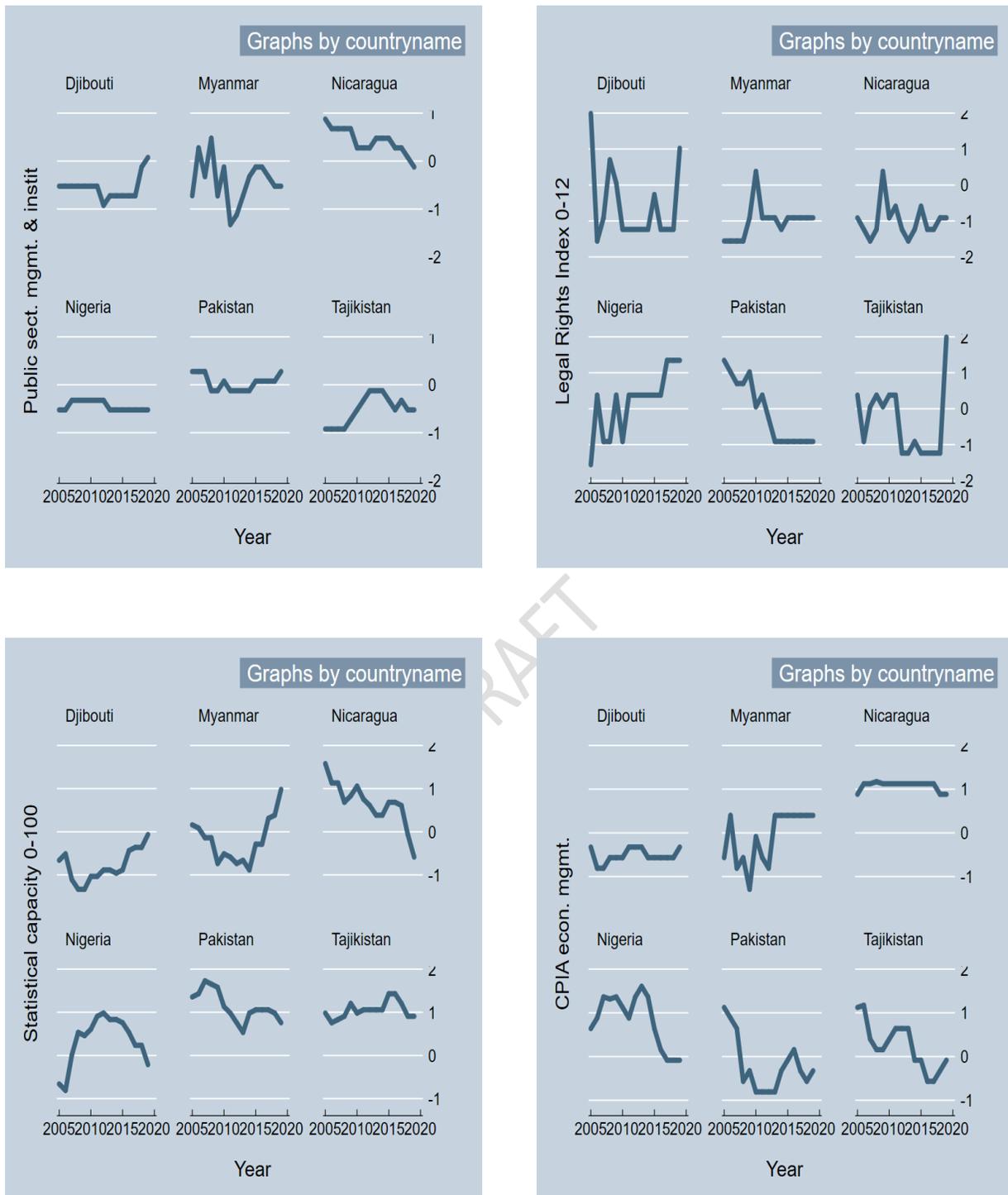

**Figures E.6: Social Capacity**

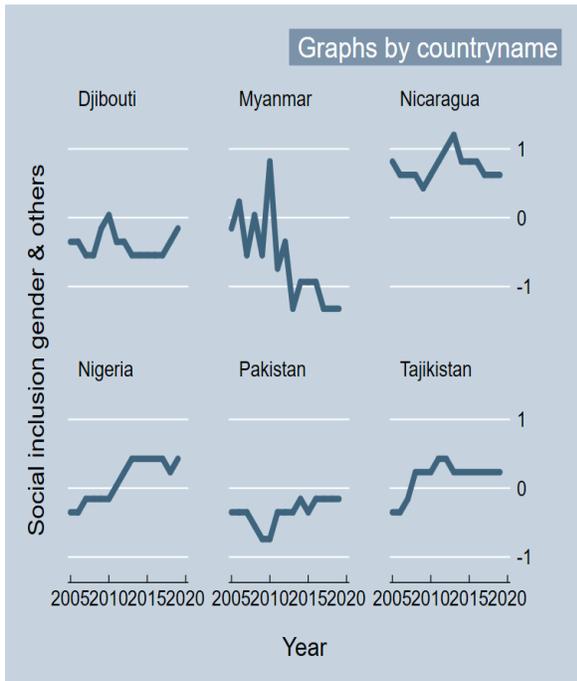

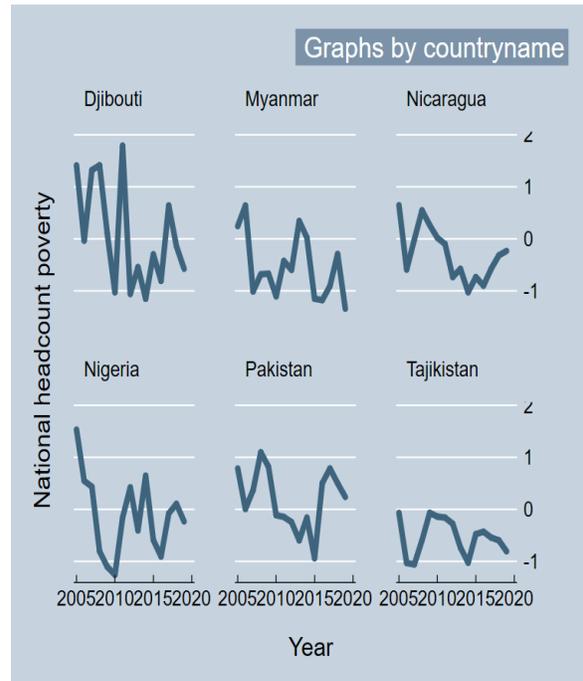

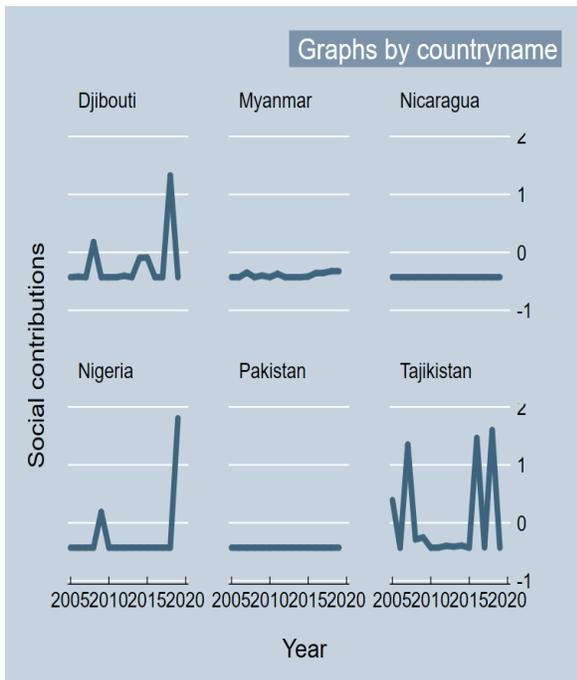

**Appendix F: Comparative Ranking of Countries According to Absorptive Capacity Index (2019).**

| Rank | Country | Tech_Index | Finance_Index | Infrastructure_Index | HumanCapacity_Index | PublicPolicy_Index | SocialCapacity_Index | AbsorptiveCapacity_Index |
|---|---|---|---|---|---|---|---|---|
| 1 | Vietnam | 1.835127 | 1.823812 | 1.932174 | 0.765487 | 1.076316 | 0.504371 | 1.322881 |
| 2 | India | 4.306017 | 0.982093 | 0.825784 | 0.428893 | -0.11115 | 0.561976 | 1.165602 |
| 3 | Bosnia and Herzegovina | 0.230912 | 0.842245 | 2.37799 | 1.014441 | 0.332861 | 1.059924 | 0.976395 |
| 4 | Kosovo | 1.237119 | 0.593937 | 1.921194 | 0.621909 | 0.626732 | 0.436277 | 0.906195 |
| 5 | Moldova | 0.387072 | 0.462939 | 2.018654 | 0.425024 | 1.032233 | 1.019524 | 0.890908 |
| 6 | Georgia | 0.378321 | 0.507115 | 2.306922 | 0.393636 | 1.286522 | 0.438898 | 0.885235 |
| 7 | Mongolia | 1.928538 | 0.183368 | 0.108906 | 1.180915 | 0.670887 | 0.814918 | 0.814589 |
| 8 | Uzbekistan | 0.811956 | -0.2022 | 1.691056 | 0.950817 | 0.510404 | 0.976198 | 0.789705 |
| 9 | Bolivia | 0.293038 | 0.894145 | 1.058183 | 1.012945 | 0.155349 | 1.158961 | 0.762103 |
| 10 | St. Vincent and the Grenadines | 1.053247 | 0.208629 | 1.541172 | 0.660106 | 0.48422 | 0.26156 | 0.701489 |
| 11 | Grenada | 0.3357 | 0.14374 | 2.150201 | 0.598516 | 0.417394 | 0.514369 | 0.69332 |
| 12 | Armenia | 0.128657 | 0.107966 | 1.320656 | 0.568944 | 1.224323 | 0.48454 | 0.639181 |
| 13 | St. Lucia | 0.461946 | 0.093167 | 1.815016 | 0.638062 | 0.351249 | 0.457002 | 0.636074 |
| 14 | Dominica | 0.132727 | 1.010012 | 1.471019 | 0.758585 | 0.462641 | -0.14163 | 0.61556 |
| 15 | Kyrgyz Republic | 0.42882 | -0.21069 | 0.918591 | 0.663019 | 0.973312 | 0.717389 | 0.581741 |
| 16 | Cabo Verde | -0.23006 | 0.358551 | 0.900362 | 0.568667 | 0.370873 | 1.104725 | 0.512186 |
| 17 | Samoa | -0.37554 | 0.426603 | 0.845797 | 0.439965 | 1.087011 | 0.596361 | 0.503366 |
| 18 | Kenya | 0.820638 | 0.242582 | 0.088082 | 0.345713 | 0.817826 | 0.390348 | 0.450865 |
| 19 | Nepal | 0.123604 | 0.321283 | 0.715523 | 0.39143 | 0.584091 | 0.486709 | 0.437107 |
| 20 | Bhutan | -0.23118 | 0.578921 | 0.435955 | 0.466053 | 0.606513 | 0.527948 | 0.397369 |
| 21 | Honduras | 0.254915 | 0.323002 | 0.605793 | 0.342447 | 0.229982 | 0.594319 | 0.391743 |
| 22 | Cambodia | 0.074192 | 0.763915 | 0.704527 | 0.409797 | 0.388614 | -0.11648 | 0.370761 |
| 23 | Sri Lanka | -0.19002 | 0.233907 | 0.967374 | 0.555663 | 0.282689 | 0.148645 | 0.333043 |
| 24 | Rwanda | 0.327286 | -0.46087 | 0.051823 | -0.06512 | 1.26666 | 0.750217 | 0.311665 |
| 25 | Nigeria | 0.4944 | -0.31568 | 0.777391 | 0.117993 | -0.00503 | 0.630286 | 0.283226 |
| 26 | Maldives | -0.45821 | 0.245947 | 1.475443 | 0.492052 | -0.15242 | 0.00429 | 0.267849 |
| 27 | Lao PDR | 0.143695 | 0.577372 | 0.567417 | 0.261909 | -0.24236 | 0.269928 | 0.262993 |
| 28 | Senegal | -0.0125 | -0.25093 | 0.39761 | -0.24882 | 0.847148 | 0.458069 | 0.198429 |
| 29 | Tonga | -0.41569 | -0.09632 | 0.519016 | 0.309611 | 0.603783 | 0.101094 | 0.170249 |
| 30 | Ghana | -0.37397 | -0.55224 | 0.604351 | 0.353584 | 0.650955 | 0.182034 | 0.144118 |
| 31 | Tanzania | 0.026761 | -0.09467 | 0.203206 | -0.48162 | 0.207781 | 0.796058 | 0.109586 |
| 32 | Cote d'Ivoire | -0.07358 | -0.22658 | 0.366575 | -0.01271 | 0.480156 | 0.108809 | 0.107113 |
| 33 | Ethiopia | 0.577417 | -0.20974 | -0.22005 | -0.11129 | 0.017556 | 0.565574 | 0.103244 |
| 34 | Djibouti | 0.225756 | -0.15364 | 0.400645 | -0.04848 | 0.154949 | -0.04464 | 0.089098 |
| 35 | Lesotho | 0.303178 | 0.38123 | 0.19675 | -0.11242 | 0.142388 | -0.403 | 0.084688 |
| 36 | Togo | -0.03731 | -0.28685 | -0.00302 | 0.256051 | 0.283281 | 0.29095 | 0.083851 |
| 37 | Bangladesh | -0.10129 | 0.4261 | 0.06743 | 0.329799 | -0.23315 | -0.04416 | 0.074122 |
| 38 | Guyana | -0.3607 | -0.37911 | 1.019161 | 0.422247 | -0.21172 | -0.0783 | 0.068595 |
| 39 | Pakistan | 0.062532 | -0.09652 | 0.137691 | 0.111673 | 0.095186 | 0.099132 | 0.068282 |
| 40 | Kiribati | 0.019126 | 0.386583 | 0.227063 | 0.653179 | -0.60597 | -0.39131 | 0.048111 |
| 41 | Vanuatu | -0.19003 | -0.04284 | 0.385256 | 0.219316 | 0.447211 | -0.55403 | 0.044149 |
| 42 | Burkina Faso | 0.135966 | -0.13823 | -0.15245 | 0.123046 | 0.317112 | -0.04392 | 0.040255 |
| 43 | Benin | -0.16928 | -0.15866 | -0.3173 | -0.00672 | 0.570117 | 0.148464 | 0.011105 |

| Rank | Country | Tech_Index | Finance_Index | Infrastructure_Index | HumanCapacity_Index | PublicPolicy_Index | SocialCapacity_Index | AbsorptiveCapacity_Index |
|------|---------|-----------|---------------|---------------------|---------------------|--------------------|--------------------|--------------------------|
| 44 | Malawi | -0.1415 | -0.29775 | -0.33264 | 0.1113 | 0.405855 | 0.154537 | -0.0167 |
| 45 | Nicaragua | -0.31277 | -0.23147 | 0.232112 | 0.103423 | -0.19638 | 0.281213 | -0.02065 |
| 46 | Tajikistan | 0.032905 | -0.56481 | 0.298435 | -0.04933 | 0.286401 | -0.15013 | -0.02442 |
| 47 | Tuvalu | 0.150786 | 0.103685 | 0.691382 | 0.070817 | -0.58147 | -0.61998 | -0.0308 |
| 48 | Uganda | -0.15365 | -0.37509 | -0.4356 | -0.30694 | 0.581907 | 0.363349 | -0.05434 |
| 49 | Gambia, The | -0.45538 | -0.08629 | 0.017895 | 0.129778 | -0.15273 | -0.02504 | -0.0953 |
| 50 | Mali | -0.21218 | -0.23106 | -0.18255 | -0.29749 | 0.386897 | -0.04213 | -0.09642 |
| 51 | Micronesia, Fed. Sts. | -0.02808 | 0.044986 | 0.069799 | 0.316725 | -0.40417 | -0.66775 | -0.11141 |
| 52 | Zambia | -0.01398 | -0.35426 | -0.04731 | -0.07948 | 0.303056 | -0.55552 | -0.12458 |
| 53 | Sao Tome and Principe | -0.06513 | 0.416852 | -0.05536 | -0.01201 | -0.52065 | -0.6106 | -0.14115 |
| 54 | Mauritania | -0.31342 | -0.27998 | -0.04341 | -0.21894 | -0.17649 | 0.098707 | -0.15559 |
| 55 | Sierra Leone | -0.04923 | -0.3396 | -0.49023 | 0.11846 | -0.16617 | -0.20295 | -0.18829 |
| 56 | Cameroon | -0.33608 | -0.41874 | 0.10723 | -0.1589 | 0.101027 | -0.43662 | -0.19035 |
| 57 | Timor-Leste | -0.37282 | -0.21221 | 0.086283 | 0.432846 | -0.67408 | -0.46295 | -0.20049 |
| 58 | Zimbabwe | -0.16584 | -0.13868 | -0.16432 | -0.48678 | -0.46139 | 0.211109 | -0.20098 |
| 59 | Myanmar | -0.29059 | 0.075565 | 0.457526 | -0.17931 | -0.12093 | -1.15453 | -0.20205 |
| 60 | Liberia | -0.47768 | -0.14699 | -0.19402 | -0.23901 | -0.16844 | -0.24811 | -0.24571 |
| 61 | Marshall Islands | 0.127603 | -0.10766 | 0.089237 | 0.330901 | -0.69479 | -1.23307 | -0.24796 |
| 62 | Niger | -0.2654 | -0.63205 | -0.42542 | -0.7227 | 0.338601 | 0.190373 | -0.25276 |
| 63 | Afghanistan | -0.2297 | 0.081822 | -0.54753 | -0.00905 | -0.39105 | -0.56674 | -0.27704 |
| 64 | Mozambique | -0.07728 | 0.071048 | -0.54299 | -0.6898 | -0.31835 | -0.26513 | -0.30375 |
| 65 | Guinea | -0.53699 | -0.69755 | -0.02544 | -0.66646 | -0.01886 | 0.105032 | -0.30671 |
| 66 | Solomon Islands | -0.02331 | -0.19956 | -0.35161 | -0.35527 | -0.07189 | -0.84626 | -0.30798 |
| 67 | Papua New Guinea | -0.42733 | -0.20056 | -0.31732 | -0.26572 | -0.01698 | -0.69641 | -0.32072 |
| 68 | Madagascar | -0.24931 | -0.22504 | -0.53249 | -0.32918 | -0.29181 | -0.31105 | -0.32314 |
| 69 | Haiti | -0.27428 | 0.586672 | -0.24086 | -0.31125 | -0.82586 | -0.91451 | -0.33001 |
| 70 | Burundi | -0.29326 | -0.5261 | -0.58438 | -0.50933 | -0.63882 | 0.326913 | -0.37083 |
| 71 | Congo, Rep. | -0.42095 | -0.33139 | -0.21451 | -0.03004 | -0.70382 | -0.61605 | -0.38613 |
| 72 | Angola | -0.53496 | -0.03026 | -0.15295 | -0.56374 | -0.83363 | -0.69364 | -0.4682 |
| 73 | Central African Republic | -0.0446 | -0.32443 | -0.57165 | -0.08778 | -0.92302 | -1.06998 | -0.50358 |
| 74 | Guinea-Bissau | -0.5636 | -0.06699 | -0.4075 | -0.51317 | -0.75815 | -0.81315 | -0.52043 |
| 75 | Comoros | -0.57087 | -0.51273 | -0.33662 | -0.39159 | -0.58277 | -0.744 | -0.5231 |
| 76 | Chad | -0.40747 | -0.30302 | -0.84945 | -0.83105 | -0.62751 | -0.23418 | -0.54211 |
| 77 | Congo, Dem. Rep. | -0.41142 | -0.67921 | -0.73101 | -0.87923 | -0.57574 | 0.008977 | -0.54461 |
| 78 | Sudan | -0.42613 | -0.42575 | 0.009877 | -0.7036 | -1.09683 | -0.84189 | -0.58072 |
| 79 | Eritrea | -0.15135 | 0.052468 | -0.34355 | 0.024908 | -2.30063 | -0.84071 | -0.59314 |
| 80 | Yemen, Rep. | 0.161517 | -0.35635 | 0.024481 | -0.53569 | -2.06751 | -1.01354 | -0.63118 |
| 81 | Somalia | -0.51377 | -0.10877 | -0.76224 | -0.49105 | -2.24777 | -0.80263 | -0.82104 |
| 82 | South Sudan | -0.66031 | -0.50391 | -0.86225 | -0.32214 | -2.27191 | -1.80079 | -1.07022 |

**Appendix G: Kernel Densities of the Observed and Complete Dataset.**

**Figures G.1: Technology capacity**

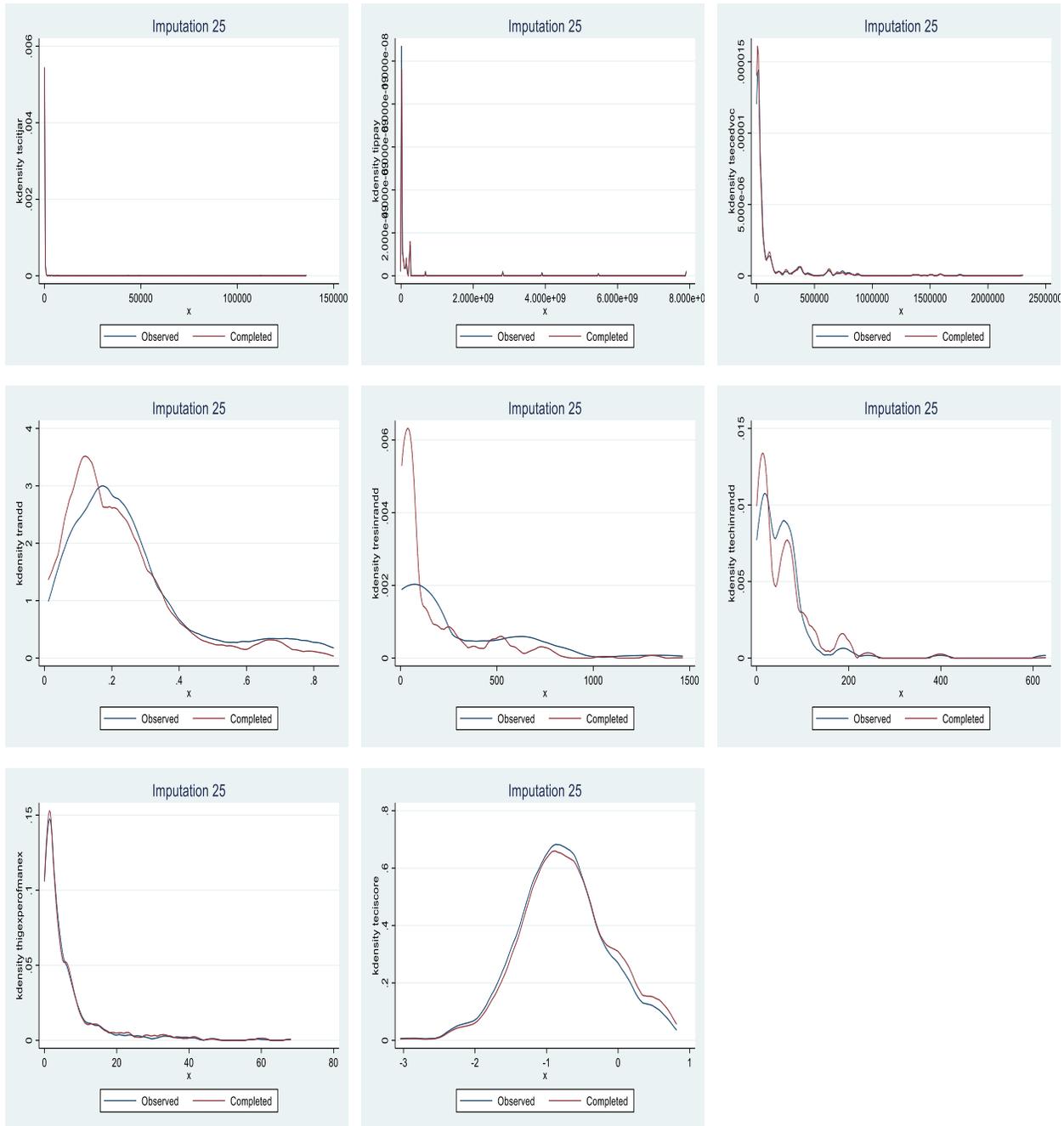

## Figures G.2: Financial Capacity

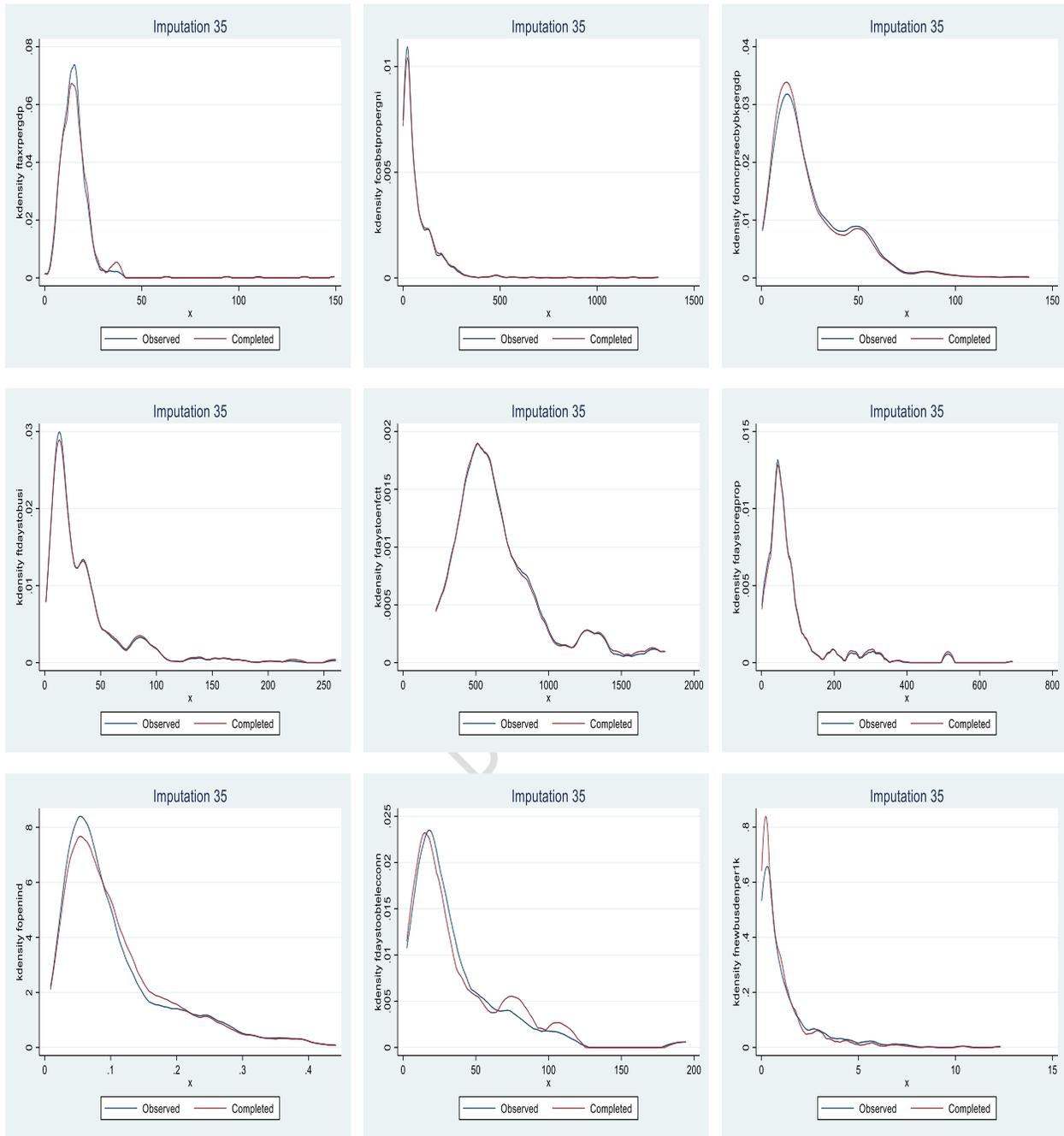

**Figures G.2: Financial Capacity (continued)**

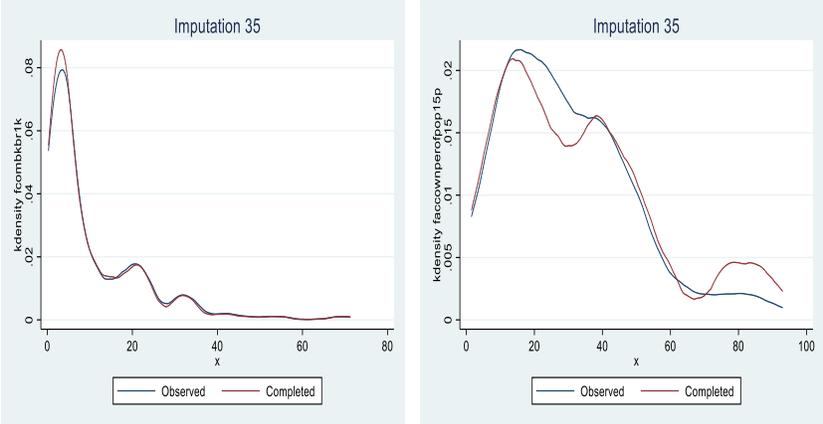

## Figure G.3: Human Capacity

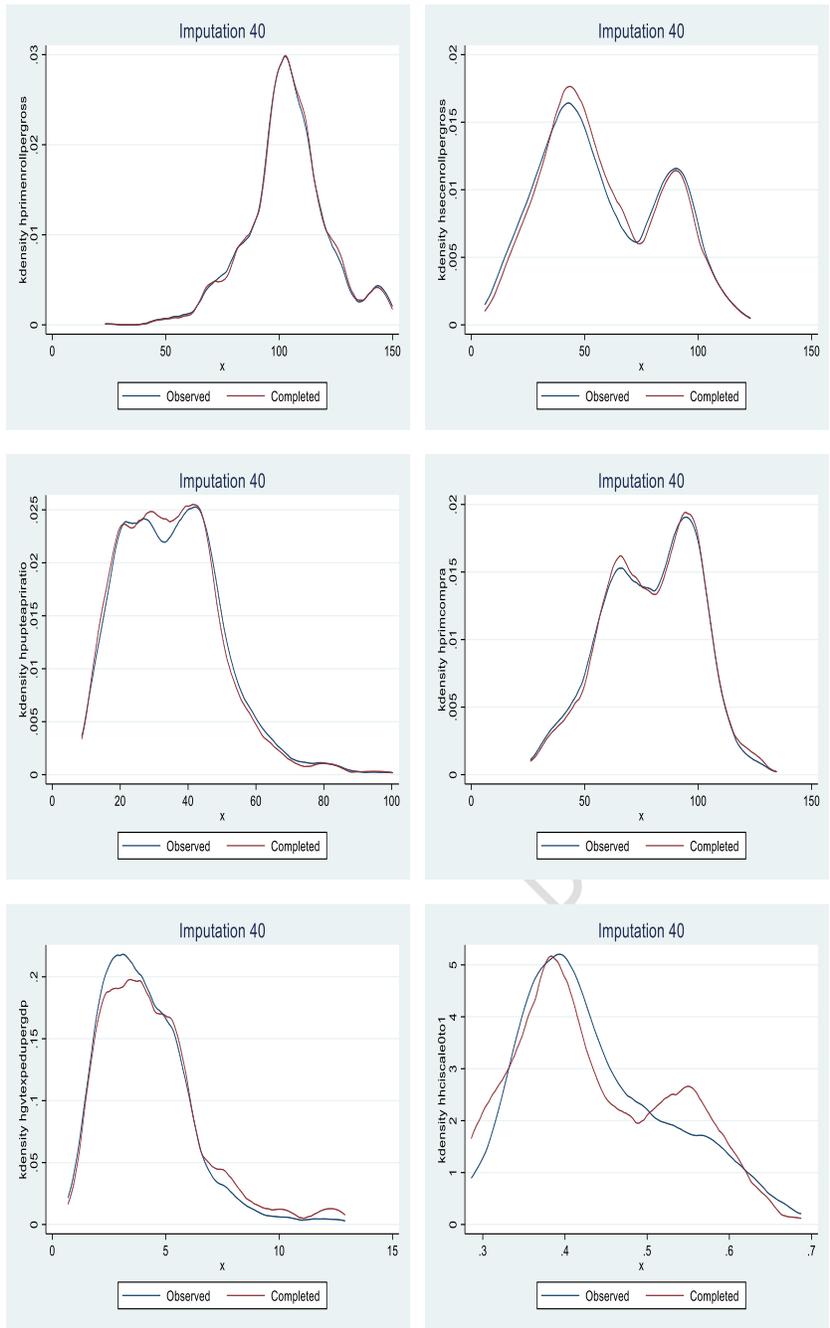

**Figure G.3: Human Capacity (Continued)**

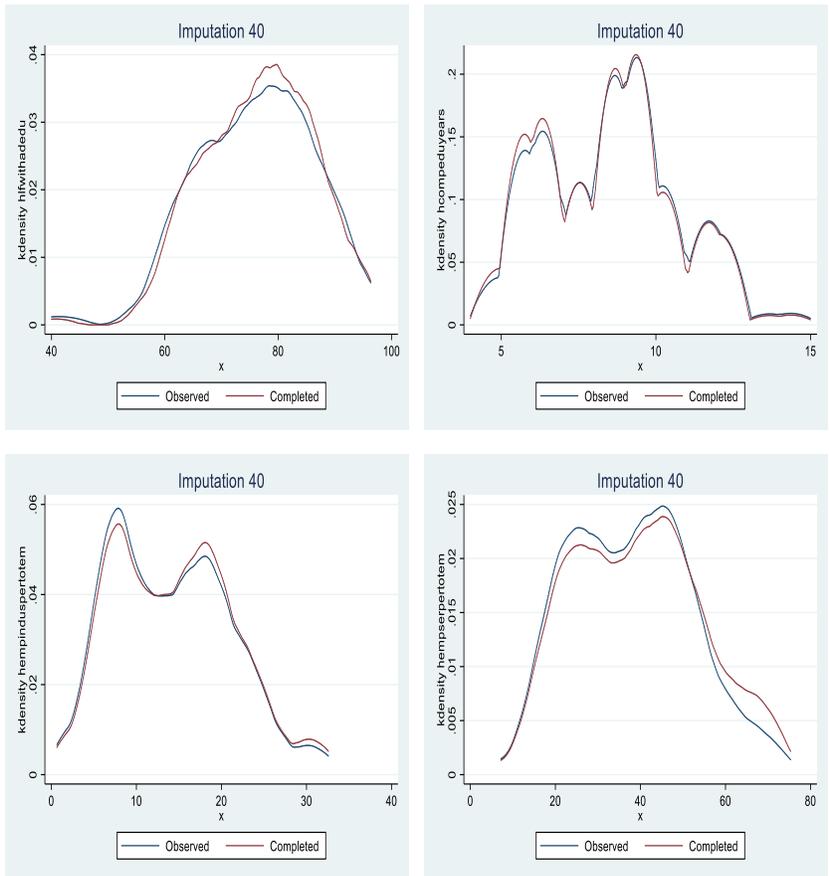

## Figures G.4: Infrastructure Capacity

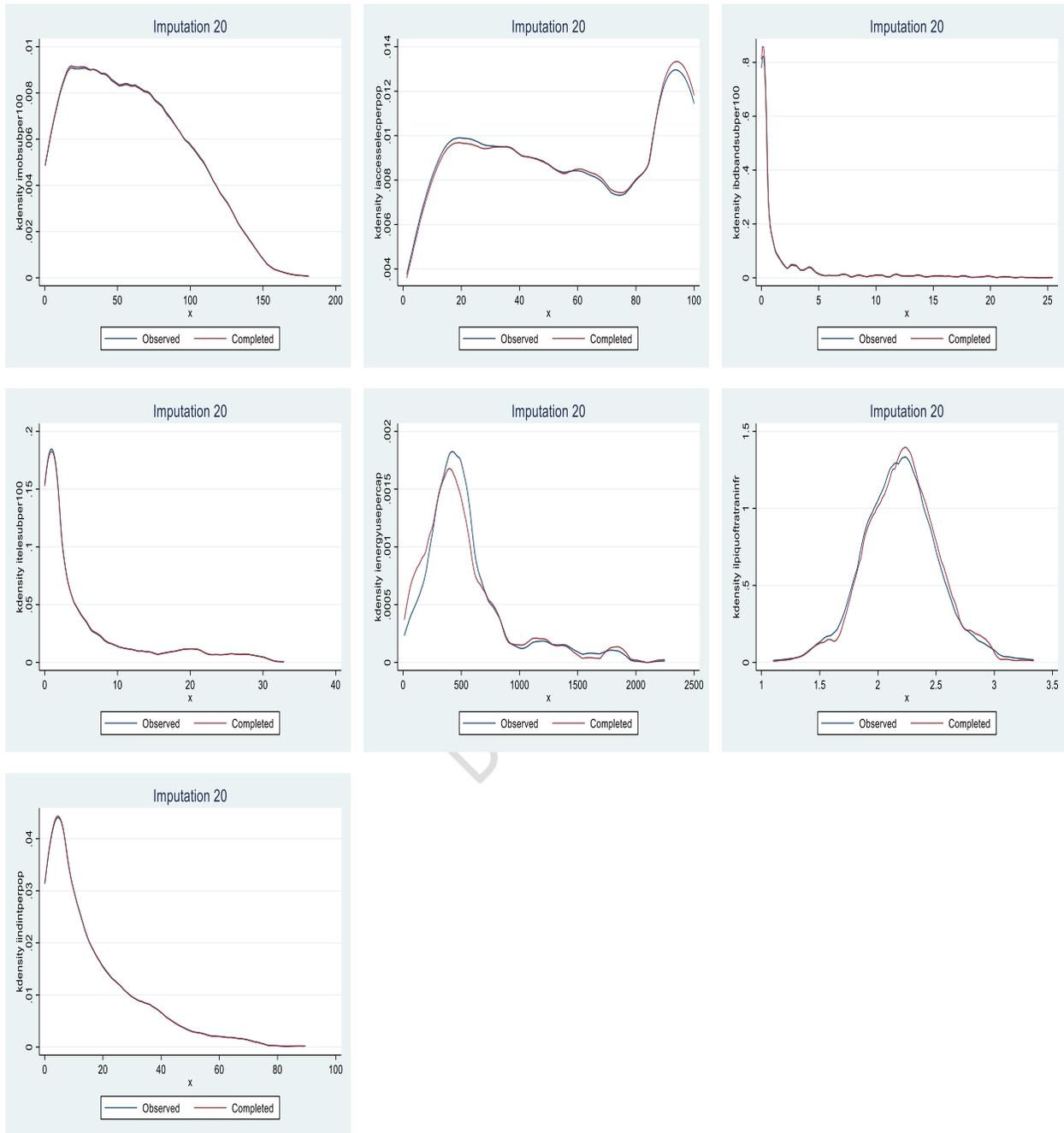

## Figures G.5: Public Policy Capacity

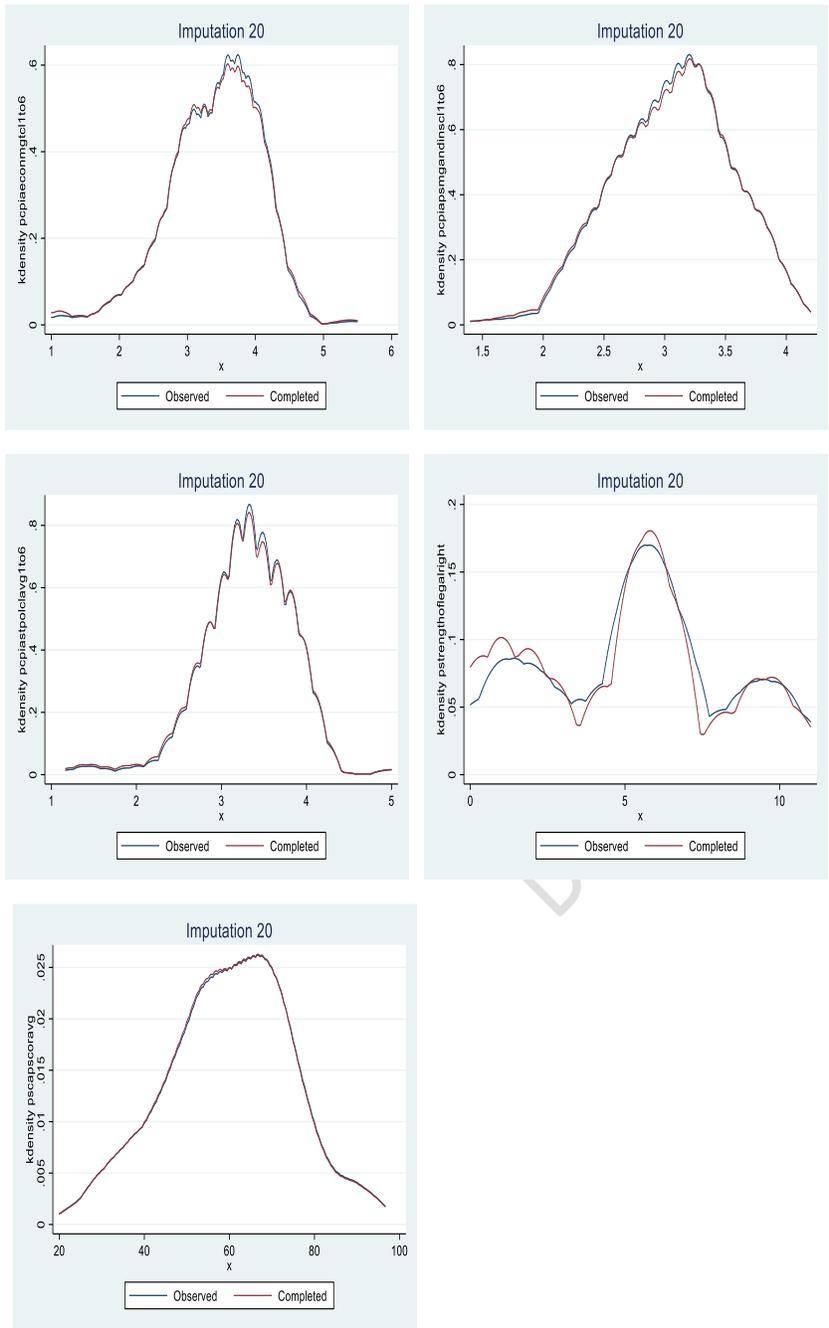

## Figures G.6: Social Capacity

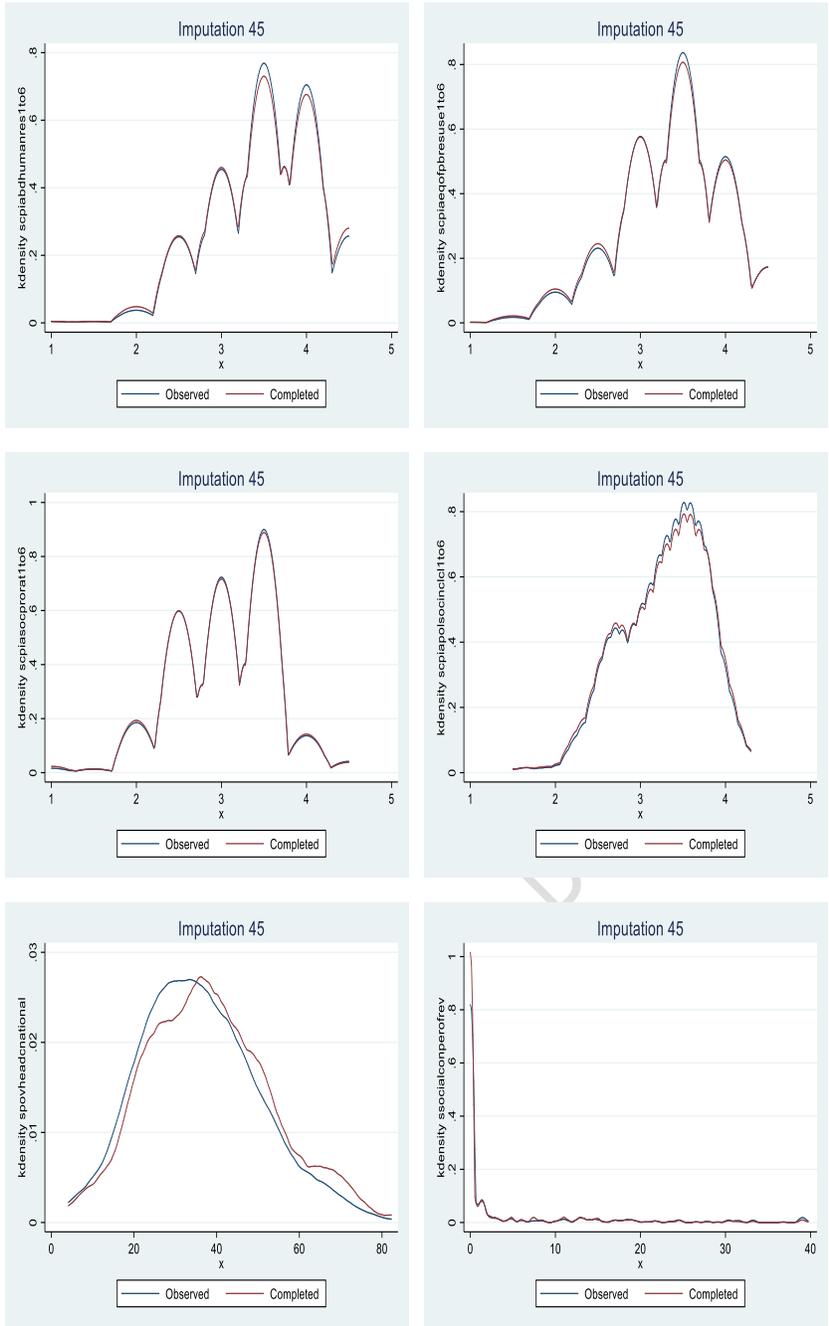

## Appendix H: Pairwise correlations for incomplete (m=0) and complete datasets (m=25)

### Tables H.1 Technology capacity pairwise correlations incomplete (above) and complete (below)

| Variables | (1) | (2) | (3) | (4) | (5) | (6) | (7) | (8) |
|---|---|---|---|---|---|---|---|---|
| **(1) Sci & tech. articles** | 1.000 | | | | | | | |
| **(2) Intellectual payments (mil)** | 0.981 | 1.000 | | | | | | |
| **(3) Voc. & tech. students (mil)** | 0.621 | 0.653 | 1.000 | | | | | |
| **(4) R&D expend. % of GDP** | 0.605 | 0.549 | 0.385 | 1.000 | | | | |
| **(5) R&D researchers (per mil)** | -0.015 | -0.020 | 0.064 | 0.187 | 1.000 | | | |
| **(6) R&D technicians (per mil)** | 0.074 | 0.071 | 0.050 | 0.244 | 0.439 | 1.000 | | |
| **(7) High-tech exports (mil)** | 0.041 | 0.061 | -0.034 | 0.170 | 0.183 | 0.013 | 1.000 | |
| **(8) ECI (econ. complexity)** | 0.244 | 0.268 | 0.096 | 0.338 | 0.545 | 0.323 | 0.080 | 1.000 |

| Variables | (1) | (2) | (3) | (4) | (5) | (6) | (7) | (8) |
|---|---|---|---|---|---|---|---|---|
| **(1) Sci & tech. articles** | 1.000 | | | | | | | |
| **(2) Intellectual payments (mil)** | 0.968 | 1.000 | | | | | | |
| **(3) Voc. & tech. students (mil)** | 0.527 | 0.487 | 1.000 | | | | | |
| **(4) R&D expend. % of GDP** | 0.346 | 0.307 | 0.264 | 1.000 | | | | |
| **(5) R&D researchers (per mil)** | -0.005 | -0.004 | 0.072 | 0.122 | 1.000 | | | |
| **(6) R&D technicians (per mil)** | 0.148 | 0.115 | 0.046 | 0.258 | 0.097 | 1.000 | | |
| **(7) High-tech exports (mil)** | 0.024 | 0.033 | -0.049 | 0.169 | 0.052 | 0.123 | 1.000 | |
| **(8) ECI (econ. complexity)** | 0.193 | 0.176 | 0.040 | 0.085 | 0.409 | 0.244 | 0.066 | 1.000 |

### Tables H.2: Financial capacity pairwise correlations incomplete (above) and complete (below)

| Variables | (1) | (2) | (3) | (4) | (5) | (6) | (7) | (8) | (9) | (10) | (11) |
|---|---|---|---|---|---|---|---|---|---|---|---|
| **(1) Tax revenue (% of GDP)** | 1.000 | | | | | | | | | | |
| **(2) Business startup cost** | -0.207 | 1.000 | | | | | | | | | |
| **(3) Domestic credit by banks** | 0.072 | -0.333 | 1.000 | | | | | | | | |
| **(4) Days to start business** | 0.127 | 0.382 | -0.134 | 1.000 | | | | | | | |
| **(5) Days enforcing contract** | 0.025 | 0.055 | -0.132 | 0.165 | 1.000 | | | | | | |
| **(6) Days to register property** | -0.045 | 0.193 | -0.166 | 0.250 | 0.199 | 1.000 | | | | | |
| **(7) Openness measure** | -0.006 | 0.077 | 0.518 | 0.552 | -0.426 | 0.071 | 1.000 | | | | |
| **(8) Days to electric meter** | -0.200 | 0.034 | -0.140 | -0.078 | 0.129 | 0.137 | -0.138 | 1.000 | | | |
| **(9) Business density** | 0.278 | -0.193 | 0.390 | -0.119 | -0.155 | -0.306 | 0.186 | -0.123 | 1.000 | | |
| **(10) Financial accountholders** | 0.171 | -0.224 | 0.400 | -0.052 | 0.020 | -0.141 | 0.134 | -0.116 | 0.475 | 1.000 | |
| **(11) Commercial banks** | 0.086 | -0.319 | 0.526 | -0.183 | -0.237 | -0.202 | 0.220 | -0.096 | 0.553 | 0.531 | 1.000 |

| Variables | (1) | (2) | (3) | (4) | (5) | (6) | (7) | (8) | (9) | (10) | (11) |
|---|---|---|---|---|---|---|---|---|---|---|---|
| **(1) Tax revenue (% of GDP)** | 1.000 | | | | | | | | | | |
| **(2) Business startup cost** | -0.068 | 1.000 | | | | | | | | | |
| **(3) Domestic credit by banks** | 0.167 | -0.270 | 1.000 | | | | | | | | |
| **(4) Days to start business** | 0.290 | 0.419 | -0.126 | 1.000 | | | | | | | |
| **(5) Days enforcing contract** | 0.175 | 0.060 | -0.068 | 0.169 | 1.000 | | | | | | |
| **(6) Days to register property** | 0.074 | 0.189 | -0.173 | 0.235 | 0.197 | 1.000 | | | | | |
| **(7) Openness measure** | 0.153 | 0.095 | 0.476 | 0.537 | -0.380 | 0.009 | 1.000 | | | | |
| **(8) Days to electric meter** | -0.131 | -0.104 | 0.082 | -0.122 | 0.081 | 0.038 | -0.081 | 1.000 | | | |
| **(9) Business density** | 0.273 | -0.145 | 0.339 | -0.125 | -0.119 | -0.201 | 0.138 | 0.026 | 1.000 | | |
| **(10) Financial accountholders** | 0.199 | -0.231 | 0.451 | -0.068 | 0.053 | -0.138 | 0.111 | 0.094 | 0.401 | 1.000 | |
| **(11) Commercial banks** | 0.043 | -0.260 | 0.501 | -0.175 | -0.216 | -0.123 | 0.218 | -0.006 | 0.455 | 0.533 | 1.000 |

**Tables H.3: Human capacity pairwise correlations incomplete (above) and complete (below)**

| Variables | (1) | (2) | (3) | (4) | (5) | (6) | (7) | (8) | (9) | (10) |
|---|---|---|---|---|---|---|---|---|---|---|
| **(1) Primary enrollment (gross)** | 1.000 | | | | | | | | | |
| **(2) Sec. enrollment (gross)** | 0.178 | 1.000 | | | | | | | | |
| **(3) Primary pupil-teacher ratio** | 0.064 | -0.787 | 1.000 | | | | | | | |
| **(4) Primary completion rate** | 0.370 | 0.867 | -0.694 | 1.000 | | | | | | |
| **(5) Govt. expend. on educ.** | 0.140 | 0.240 | -0.261 | 0.252 | 1.000 | | | | | |
| **(6) Human Capital Index 0-1** | 0.052 | 0.908 | -0.717 | 0.792 | 0.164 | 1.000 | | | | |
| **(7) Advanced educ. labor** | 0.171 | -0.005 | 0.005 | 0.001 | -0.143 | 0.251 | 1.000 | | | |
| **(8) Compulsory educ. (years)** | -0.288 | 0.338 | -0.260 | 0.171 | 0.234 | 0.364 | -0.126 | 1.000 | | |
| **(9) Industry employment** | -0.044 | 0.637 | -0.546 | 0.538 | 0.060 | 0.534 | 0.001 | 0.306 | 1.000 | |
| **(10) Service employment** | -0.162 | 0.620 | -0.648 | 0.449 | 0.222 | 0.372 | -0.109 | 0.268 | 0.559 | 1.000 |

| Variables | (1) | (2) | (3) | (4) | (5) | (6) | (7) | (8) | (9) | (10) |
|---|---|---|---|---|---|---|---|---|---|---|
| **(1) Primary enrollment (gross)** | 1.000 | | | | | | | | | |
| **(2) Sec. enrollment (gross)** | 0.174 | 1.000 | | | | | | | | |
| **(3) Primary pupil-teacher ratio** | 0.020 | -0.708 | 1.000 | | | | | | | |
| **(4) Primary completion rate** | 0.372 | 0.815 | -0.646 | 1.000 | | | | | | |
| **(5) Govt. expend. on educ.** | 0.107 | 0.325 | -0.284 | 0.346 | 1.000 | | | | | |
| **(6) Human Capital Index 0-1** | 0.187 | 0.796 | -0.619 | 0.723 | 0.204 | 1.000 | | | | |
| **(7) Advanced educ. labor** | 0.011 | -0.129 | 0.181 | -0.144 | -0.067 | -0.034 | 1.000 | | | |
| **(8) Compulsory educ. (years)** | -0.306 | 0.335 | -0.211 | 0.178 | 0.176 | 0.308 | -0.076 | 1.000 | | |
| **(9) Industry employment** | -0.024 | 0.633 | -0.529 | 0.514 | 0.147 | 0.494 | -0.174 | 0.345 | 1.000 | |
| **(10) Service employment** | -0.105 | 0.623 | -0.641 | 0.446 | 0.263 | 0.472 | -0.163 | 0.313 | 0.565 | 1.000 |

**Tables H.4: Infrastructure capacity pairwise correlations incomplete (above) and complete (below)**

| Variables | (1) | (2) | (3) | (4) | (5) | (6) | (7) |
|---|---|---|---|---|---|---|---|
| **(1) Mobile subscriptions** | 1.000 | | | | | | |
| **(2) Access to electricity** | 0.514 | 1.000 | | | | | |
| **(3) Broadband subscriptions** | 0.490 | 0.519 | 1.000 | | | | |
| **(4) Telephone subscriptions** | 0.343 | 0.682 | 0.694 | 1.000 | | | |
| **(5) Energy use (per capita)** | 0.371 | 0.567 | 0.573 | 0.556 | 1.000 | | |
| **(6) Logistic perf. Index 1-5** | 0.344 | 0.250 | 0.244 | 0.160 | 0.154 | 1.000 | |
| **(7) Internet users** | 0.680 | 0.651 | 0.733 | 0.579 | 0.580 | 0.343 | 1.000 |

| Variables | (1) | (2) | (3) | (4) | (5) | (6) | (7) |
|---|---|---|---|---|---|---|---|
| **(1) Mobile subscriptions** | 1.000 | | | | | | |
| **(2) Access to electricity** | 0.509 | 1.000 | | | | | |
| **(3) Broadband subscriptions** | 0.471 | 0.496 | 1.000 | | | | |
| **(4) Telephone subscriptions** | 0.342 | 0.664 | 0.684 | 1.000 | | | |
| **(5) Energy use (per capita)** | 0.363 | 0.559 | 0.702 | 0.585 | 1.000 | | |
| **(6) Logistic perf. Index 1-5** | 0.238 | 0.261 | 0.100 | 0.092 | 0.115 | 1.000 | |
| **(7) Internet users** | 0.669 | 0.643 | 0.732 | 0.571 | 0.592 | 0.240 | 1.000 |

**Tables H.5: Public Policy capacity pairwise correlations incomplete (above) and complete (below)**

| Variables | (1) | (2) | (3) | (4) | (5) |
|---|---|---|---|---|---|
| **(1) CPIA econ. mgmt.** | 1.000 | | | | |
| **(2) Public sect. mgmt. & instit** | 0.612 | 1.000 | | | |
| **(3) Structural policies** | 0.649 | 0.740 | 1.000 | | |
| **(4) Statistical capacity 0-100** | 0.498 | 0.437 | 0.527 | 1.000 | |
| **(5) Legal Rights Index 0-12** | 0.218 | 0.189 | 0.293 | 0.067 | 1.000 |

| Variables | (1) | (2) | (3) | (4) | (5) |
|---|---|---|---|---|---|
| **(1) CPIA econ. mgmt.** | 1.000 | | | | |
| **(2) Public sect. mgmt. & instit** | 0.625 | 1.000 | | | |
| **(3) Structural policies** | 0.641 | 0.740 | 1.000 | | |
| **(4) Statistical capacity 0-100** | 0.518 | 0.493 | 0.558 | 1.000 | |
| **(5) Legal Rights Index 0-12** | 0.182 | 0.274 | 0.337 | 0.160 | 1.000 |

**Tables H.6: Social capacity pairwise correlations incomplete (above) and complete (below)**

| Variables | (1) | (2) | (3) | (4) | (5) | (6) |
|---|---|---|---|---|---|---|
| **(1) Human resources rating** | 1.000 | | | | | |
| **(2) Equity of public resc use** | 0.620 | 1.000 | | | | |
| **(3) Social protection rating** | 0.627 | 0.655 | 1.000 | | | |
| **(4) Social inclusion o..** | 0.852 | 0.827 | 0.815 | 1.000 | | |
| **(5) National headcount poverty** | -0.387 | -0.244 | -0.374 | -0.410 | 1.000 | |
| **(6) Social contributions** | 0.213 | 0.169 | 0.326 | 0.338 | -0.187 | 1.000 |

| Variables | (1) | (2) | (3) | (4) | (5) | (6) |
|---|---|---|---|---|---|---|
| **(1) Human resources rating** | 1.000 | | | | | |
| **(2) Equity of public resc use** | 0.640 | 1.000 | | | | |
| **(3) Social protection rating** | 0.644 | 0.658 | 1.000 | | | |
| **(4) Social inclusion o..** | 0.865 | 0.832 | 0.819 | 1.000 | | |
| **(5) National headcount poverty** | -0.395 | -0.225 | -0.303 | -0.364 | 1.000 | |
| **(6) Social contributions** | 0.217 | 0.131 | 0.290 | 0.305 | -0.156 | 1.000 |

**Appendix I:  Checking for Convergence through Trace Plots**.

Trace plots show that the mean and standard deviations from 30 chains are converging for the imputed values.

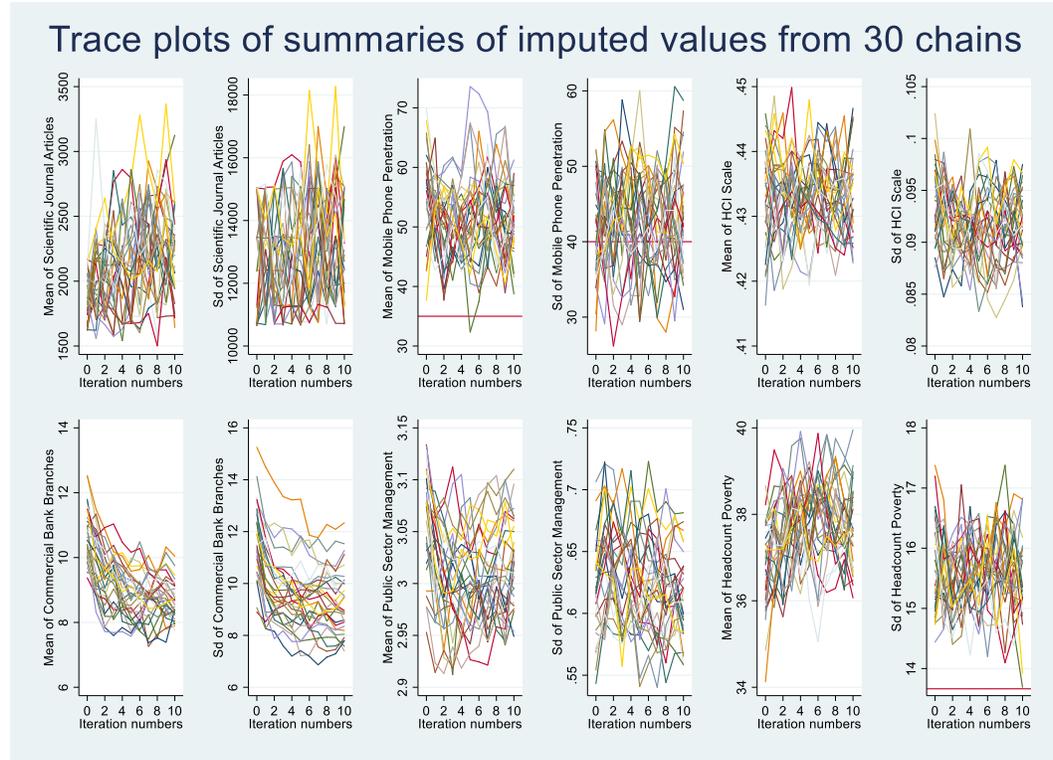